\documentclass[%
 reprint,
 longbibliography,
 amsmath,amssymb,
 aps,
]{revtex4-2}

\usepackage{graphicx}
\usepackage{dcolumn}
\usepackage{bm}

\usepackage[T1]{fontenc}
\begin{document}

\preprint{APS/123-QED}

\title{Exploring the role of accretion shocks in galaxy clusters \\ as sources of ultrahigh-energy cosmic rays}

\author{A.D. Supanitsky}
\email{daniel.supanitsky@iteda.gob.ar}
\affiliation{Instituto de Tecnologías en Detección y Astropartículas (ITeDA, CNEA-CONICET-UNSAM),
Centro Atómico Constituyentes, PC 1650, San Martín, Buenos Aires, Argentina\\}
 
\author{S.E. Nuza}
\email{snuza@iafe.uba.ar}
\affiliation{Instituto de Astronomía y Física del Espacio (IAFE, CONICET-UBA),
PC 1428 Buenos Aires, Argentina
}

\date{\today}

\begin{abstract}

Recently, the Pierre Auger Observatory has found strong evidence supporting the extragalactic origin of the most energetic cosmic 
rays. Despite several observed excesses in the distribution of arrival directions for the highest energy cosmic rays, the sources
remain unidentified. Accretion shocks in galaxy clusters have been proposed as potential sources in the past. These immense shock
waves, which can have radii on the order of megaparsecs, are generated by the infall of material from the intergalactic medium 
into the gravitational potential wells of galaxy clusters. In this work, we investigate the possibility that ultrahigh-energy 
cosmic rays are accelerated in these regions. Nearby massive galaxy clusters, including Virgo, are treated as a 
discrete component of the cluster mass distribution. Less massive galaxy clusters, as well as distant massive ones, are assumed 
to follow a continuous distribution in agreement with cluster mass statistics. We fit the flux at Earth and the composition 
profile measured by the Pierre Auger Observatory, assuming the injection of different nuclear species by these sources, to 
determine the values of the model parameters. Our results indicate that cosmic ray acceleration in cluster accretion shocks may 
account for at least a fraction of the observed UHECR flux at energies below the suppression scale. At higher energies, direct 
acceleration from the thermal pool would be feasible only if local fluctuations create favorable conditions, such as magnetic 
fields about an order of magnitude stronger than those typically expected in cluster accretion shocks, or for particular shock 
normal-magnetic field configurations. Moreover, in these scenarios, where the energy spectrum of accelerated cosmic rays is 
modeled as $\propto E^{-\gamma}$ with an exponential cutoff, the spectral index obtained from fitting the experimental data lies 
in the range $\gamma\sim 0.7-1.6$. This is smaller than $2$, which is the value expected from the first-order Fermi acceleration 
mechanism for strong shocks.

\end{abstract}

\maketitle


\section{\label{sec:Int} Introduction}

The origin of ultrahigh-energy cosmic rays (UHECRs, $E\geq 10^{18}$ eV) remains an open question 
in high-energy astrophysics. However, the vast amount of data collected by current observatories 
has led to the discovery of several key features of these highly energetic particles, which are 
crucial for understanding their nature. The observatories currently in operation include the 
Pierre Auger Observatory (Auger) \cite{AugerObs:15} in the southern hemisphere and Telescope 
Array (TA) \cite{TA:08} in the northern hemisphere, Auger being the bigger one covering an area 
of $\sim 3{\small,}000$ km$^2$. 

The energy spectrum of the UHECRs has been measured with great accuracy, thanks to the large 
exposure accumulated, particularly by Auger. Several distinctive features are observed: a hardening 
at approximately $5\times 10^{18}$ eV, known as the ankle, a steepening at around $10^{19}$ eV 
referred to as the instep; and a sharp drop at approximately $5 \times 10^{19}$ eV known as the 
suppression \cite{AugerFluxUL:20,AugerFlux:21}. Although there are various models explaining the 
origin of these features, they remain the subject of ongoing debate. 

The composition of the UHECRs is inferred mainly from the atmospheric depth of the shower maximum,
$X_\textrm{max}$, a parameter very sensitive to the primary mass together with the muon content of
the showers \cite{Supanitsky:22}. $X_\textrm{max}$ is measured with the fluorescence telescopes, 
which take data on clear and moonless nights. Therefore, the duty cycle is smaller compared to that 
of the surface detectors, reducing the statistics of events with available $X_\textrm{max}$ by 
approximately a factor of 10. Composition analyses rely on comparing experimental data with simulations 
of atmospheric air showers. Since the high-energy hadronic interactions at ultrahigh energies are not 
fully understood, air shower simulations use models that extrapolate low-energy accelerator data to 
these higher energies. This approach introduces non-negligible systematic uncertainties in the 
composition determination. Despite these challenges, the Auger $X_\textrm{max}$ data indicate that 
the composition appears to become progressively lighter from energies below $10^{18}$ eV up to around 
$10^{18.3}$ eV, where a change in composition occurs \cite{AugerXmax:19,AugerCompo:23}. Above this 
energy, the composition seems to become increasingly heavier. Additionally, the variance of 
$X_\textrm{max}$ suggests that the composition is becoming purer as the energy increases. Recently, 
Auger significantly increased the statistics of $X_\textrm{max}$ data by approximately a factor of 
$10$ after incorporating surface detector measurements in the analysis \cite{AugerXmax:25a,AugerXmax:25b}. 
This enhanced dataset reveals evidence of three distinct breaks in the mean value of $X_\textrm{max}$ as
a function of the logarithm of energy. These features are observed near the ankle, the instep, and the
suppression regions. Furthermore, the transition toward a heavier primary composition is confirmed up to 
the suppression region, a result made possible by the substantial improvement in statistics.

Regarding the distribution of the arrival directions, Auger has measured a dipolar anisotropy above 
$8\times 10^{18}$ eV \cite{AugerAniso:23,AugerDipole:17}. The amplitude of the dipole is approximately
$7.3\%$, with the significance of the measurements reaching $6.9\, \sigma$. One of the most important 
implications of this finding is that a galactic origin of the UHECRs seems to be very unlikely. At small 
and intermediate angular scales several excesses have been found by both, Auger and TA observatories. In 
the case of Auger, the data show an excess in the region of the radio galaxy Centaurus A but the significance 
of this excess is at the level of $4\, \sigma$ \cite{AugerAniso:23}. Similarly, data from the TA observatory 
show two excesses \cite{TAAniso:24}. The first is known as the hotspot, and it is located near the Ursa 
Major constellation, with the excess showing a significance level of  $2.8\, \sigma$. The second one is 
located in the region of Perseus-Pisces supercluster, with the excess showing a significance of $3.3\, \sigma$. 
It is worth mentioning that Auger analyzed these two sky regions with comparable statistics and found no 
significant signal, indicating that more studies have to be performed to understand this tension. Moreover, 
Auger data also reveal an excess in the directions of several nearby starburst galaxies, with a significance 
of $3.8\, \sigma$ \cite{AugerAniso:23}. Therefore, despite the great effort made, the sources of UHECRs have 
not yet been identified.

The origin of the ankle remains unclear, and two main models have been proposed to explain it. The first 
suggests the existence of two populations of sources: one responsible for the lighter cosmic rays (CRs) below
the ankle, and the other, consisting of intermediate and possibly heavy nuclear species, becoming significant
at energies above the ankle \cite{Aloisio:14,Mollerach:20,AugerFit:23}. In the second scenario, the light component 
below the ankle is produced by the photodisintegration of high-energy heavy nuclei in the photon field of the 
sources or their environment \cite{Unger:15,Globus:15,Globus:15b,Kach:17,Fang:18,Supanitsky:18}. The origin of 
the instep and the suppression are still being investigated. These two features appear to result from the 
exponential cutoff of the injected spectrum, modified by the propagation of the helium and nitrogen components 
in the intergalactic medium, respectively \cite{AugerFit:23}. 

Accretion shocks in galaxy clusters have also been suggested as potential sources of UHECRs 
\cite{Kang:97,Inoue:07,Ptuskin:19,Simeon:23,Blandford:23,Blandford:25}. These giant accretion shock waves can 
span several megaparsecs in radius and are capable of accelerating particles to the highest energies. They are 
generated by the infall of material from the intergalactic medium onto the high-density cluster regions within 
the cosmic web. In Ref.~\cite{Kang:97}, UHECRs from galaxy cluster accretion shocks are modeled considering only 
protons for their composition, in line with expectations before the advent of Auger. However, both observations 
and cosmological simulations suggest the presence of metals in galaxy cluster outskirts (see e.g. Ref.~\cite{Li2023}). 
In Ref.~\cite{Inoue:07}, heavier nuclei are also considered assuming a continuous distribution of sources, which 
fails to properly account for the contribution of nearby sources, an important factor in this case, especially due 
to the presence of the Virgo supercluster. In Ref.~\cite{Ptuskin:19}, only the contribution of the Virgo cluster 
is studied, without considering other nearby or more distant galaxy clusters. Finally, in 
Refs.~\cite{Simeon:23,Blandford:23}, the authors discuss the possibility that UHECRs are accelerated in galaxy 
cluster accretion shocks without fitting the measured energy spectrum and composition profile. In particular, 
in Refs.~\cite{Blandford:23,Blandford:25}, a hierarchical model in which CRs of all energies are explained through 
the action of the diffusive shock acceleration (DSA) mechanism across progressively larger scales is proposed. In 
this scenario, a high-energy population of CRs accelerated in the termination shocks of powerful galaxies can undergo 
further acceleration in the accretion shocks of galaxy clusters,¸ leading to the UHECR component.

In this work, we develop a model for the highest-energy component of UHECRs (the one above the ankle), where 
these  extremely energetic particles are accelerated in the accretion shocks of galaxy clusters. The model 
assumes that the sources inject a mixture of nuclear species, including light, intermediate, and heavy nuclei. 
We assume that CRs are accelerated via the first-order Fermi mechanism. To estimate two key parameters required 
by DSA, i.e. the magnetic field in the acceleration region and the velocity of the shock waves, we use two different 
approaches: the first is based on simplified physical considerations, while the second additionally relies on 
cosmological simulations. The contribution of nearby massive galaxy clusters is taken into account considering a 
discrete set of nearby sources, including Virgo. The contribution of less massive galaxy clusters in our vicinity, 
and of more distant ones, are also included in our model assuming a continuous distribution of sources following the 
universal halo mass function determined from cosmological simulations. We determine model parameters by fitting the 
Auger spectrum and composition profile and, based on the results of this fit, discuss the possibility that accretion 
shocks in galaxy clusters are responsible for the observed UHECRs.

This article is organized as follows. In Sec.~\ref{sec:IS}, we describe the model for the injected spectrum of 
the sources including the estimation of the magnetic field in the outskirts of galaxy clusters, the velocity of 
the infalling shock waves and the maximum energy attained by the CRs, which is obtained from the acceleration and 
interaction times of particles in the acceleration region. Also included in this section is a description of the 
discrete sample of local clusters considered. In Sec.~\ref{sec:Charac}, we present the method used to calculate 
the different contributions to the spectrum and composition profile, which is based on the simulation of the 
propagation of CRs in the intergalactic medium from the acceleration site to the observer. We also describe the fit 
procedure and the results obtained. In Sec.~\ref{sec:Disc}, we discuss our results. Finally, in Sec.~\ref{sec:Conc}, 
we present our conclusions. 

Throughout this paper, we adopt a set of cosmological parameters consistent with the Planck cosmology given by 
Ref.~\cite{PlanckCosmology14}. Specifically, we assume a total matter density of $\Omega_{\rm M} = 0.315$, a 
cosmological constant density of $\Omega_{\Lambda} = 0.685$, and a reduced Hubble constant of $h=0.673$.

\section{\label{sec:IS} Injected spectrum}

\subsection{Cosmic ray accelerators}

In this section, CR acceleration within large-scale accretion shocks developed in the external regions of galaxy clusters is modeled. CR protons 
and nuclei are assumed to gain energy by multiple shock crossings until they escape the accelerator region to further propagate through the 
intergalactic medium. These locations are therefore considered as plausible sites for CR acceleration that may be responsible for, at least, 
some of the UHECR observed population. One key ingredient in the modeling of accretion shocks is the galaxy cluster mass. Since the total 
gravitating mass of a system is better represented by the so-called {\it virial} mass, $M_{\rm vir}$ (i.e., the total mass within the virial 
radius, $R_{\rm vir}$), virial quantities will be used to characterize galaxy clusters, unless stated otherwise. The virial radius corresponds 
to the distance from the center of the cluster for which the enclosing mean mass density is equal to the virial density contrast, 
$\Delta_{\rm vir}$, times the critical density of the Universe, where $\Delta_{\rm vir}=\Delta_{\rm vir}(\Omega_i,z)$ depends both on cosmology 
and redshift \cite{BN98}.  In the following, the main ingredients of the model concerning magnetic fields and shocked gas velocities at the 
acceleration sites are presented.

\subsubsection{Magnetic fields at cluster outskirts}
\label{sec:B_acc_sh}

The magnetic field $B$ at the location of accretion shocks is computed assuming that the thermal pressure of the gas is proportional to the 
magnetic energy density. Therefore, 
\begin{equation}
\frac{B^2}{8\pi}=\beta^{-1}\frac{\rho}{\mu\, m_{\rm p}}k T,
\end{equation}

\noindent where $\beta$ is the plasma beta parameter, $\rho$ and $T$ are the gas density and temperature in the shock region, $k$ is the 
Boltzmann constant, $\mu$ is the mean molecular weight and $m_{\rm p}$ is the proton mass. Solving for the magnetic field the following 
expression is obtained
\begin{equation}
B(P)=200\,\mu{\rm G}\times\sqrt{\beta^{-1}\,P\,{\rm keV}^{-1}\,{\rm cm}^3}. 
\label{eq:Bmag}
\end{equation}

In this equation, $P$ is the gas pressure at the distance of the accretion shock. In this way, the magnetic field strength can be directly 
related to the gas pressure, thereby encapsulating the dependence on cluster mass and redshift through the pressure at the acceleration 
site. To account for the latter in galaxy clusters, the validity of the universal pressure profile determined in Ref.~\cite{Arnaud10} is 
assumed, based on the REXCESS survey \cite{Boehringer07}, a representative sample of 33 local galaxy clusters up to $z=0.2$. This function 
parametrizes the gas pressure profile for radial distances from the cluster center at $0.03\leq r/R_{500} \leq 4$, where $R_{500}$ is the 
radius enclosing a mass density 500 times the critical density of the Universe, as a function of $M_{500}\equiv M_{\rm tot}(r\leq R_{500})$ 
and redshift $z$. Since the cluster virial radius is $R_{\rm vir}\sim 2R_{500}$, this profile is thus generally valid at cluster outskirts, 
at least out to $r\sim 2 R_{\rm vir}$. This is important because accretion shocks surrounding massive galaxy clusters can be formed in the 
cluster outskirts, up to a distance of a few times the virial radius, as shown, for instance, in Ref.~\cite{Rost24}. 

Once the ambient gas pressure, $P_1$, at the location of external shocks is computed using the universal pressure profile from 
Ref.~\cite{Arnaud10}, the pressure in the downstream region, $P_2$, can be determined using the Rankine-Hugoniot conditions for 
hydrodynamical shocks. The pressure jump between the pre- and postshock regions is (e.g. Ref.~\cite{Landau59}) 
\begin{equation}
\frac{P_1}{P_2}=\frac{(\gamma_\textrm{a}+1)-(\gamma_\textrm{a}-1)\chi}{(\gamma_\textrm{a}+1)\chi-(\gamma_\textrm{a}-1)}, 
\end{equation}

\noindent where $\chi$ is the compression ratio that depends on the shock Mach number $\mathcal{M}$ and $\gamma_\textrm{a}$ is the adiabatic 
index. To compute the pressure jump in the accretion shocks, it is necessary to provide their Mach number. One possibility is to estimate 
its value based on an approximation for the infall shock velocity and the local sound speed of the intracluster medium. However, to avoid 
making assumptions about arbitrary shock velocities, the Mach number {\it stacked} profile, $\mathcal{M}(r)$, of the material surrounding 
massive galaxy clusters is directly employed, as provided by the cosmological simulations of Ref.~\cite{Rost24} (see, e.g., their Fig.~6), 
where typical average Mach number values at these locations are approximately $\mathcal{M} \gtrsim 10$. Therefore, by computing 
$B_2 \equiv B(P_2)$ in Eq.~(\ref{eq:Bmag}) it is possible to obtain a magnetic field estimate in the shock region.   

As shown in Ref.~\cite{Ha23}, accretion shocks in the outskirts of galaxy clusters exhibit a plasma beta parameter of $\beta \gtrsim 100$. 
This is in contrast with the case of internal shocks which typically display lower values. In the calculations below, a value of $\beta=100$ is 
assumed, unless otherwise stated. This yields magnitudes below the $\mu$G level in agreement with expectations from numerical simulations.

\subsubsection{Accretion shock velocity}

The material falling onto the cluster potential wells get accelerated to a terminal velocity $v_{\rm sh}$. The infall velocity of 
noncollisional particles onto a galaxy cluster of mass $M_{\rm vir}$ at a distance $r_{\rm sh}$ from its center is
\begin{equation}
v_{\rm sh,in}(r_{\rm sh})=\sqrt{\frac{2 G M_{\rm vir}}{r_{\rm sh}}},
\label{eq:vsh_in}
\end{equation}

\noindent where $G$ is Newton's gravitational constant. Since external shocks in galaxy clusters are usually generated at distances of 
$r_{\rm sh}\gtrsim R_{\rm vir}$, the infall shock velocity follows a virial mass scaling of $v_{\rm sh,in}\propto M_{\rm vir}^{1/3}$.

This approximation, however, does not take into account the hydrodynamical nature of the infalling gas. Therefore, an additional estimate, 
dubbed $v_{\rm sh,sim}$, is performed by adopting the universal cluster temperature profile, $T(r)$, given in Ref.~\cite{Loken02} as a 
function of cluster virial temperature, $T_{\rm vir}$, to get a measure of the sound speed of the intracluster medium as a function of 
cluster radius. By combining the radial function of the sound speed with the mass-dependent Mach number profile of Ref.~\cite{Rost24}, 
shock velocities in the external cluster regions are computed using the definition of Mach number, namely
\begin{equation}
v_{\rm sh,sim}(r_{\rm sh})=\mathcal{M}(r_{\rm sh}) c_{\rm s}(r_{\rm sh}), 
\label{eq:vsh_sim}
\end{equation}

\noindent where $\mathcal{M}(r_{\rm sh})$ and $c_{\rm s}(r_{\rm sh})$ are the assumed Mach number average profile and the sound speed, 
respectively, evaluated at the location of the shock front.

\subsection{Accretion shock models}

The approximations discussed in the previous section allow for the assumption of plausible astrophysical scenarios for accretion shocks 
in galaxy clusters, based on their behavior in the $B-v_{\rm sh}$ plane, which will be used to calculate the maximum cosmic-ray acceleration 
velocities in Sec.~\ref{sec:MaxEn}. In this paper, four different physically motivated models for the magnetic field and infall shock 
velocity scalings as a function of cluster virial mass and redshift are considered. In the first scenario (model A), scalings are built from 
an array of cluster masses using the infall shock velocity given by Eq.~(\ref{eq:vsh_in}) and the magnetic field $B_2$ evaluated at the virial 
radius. 
\begin{table}[t!h]
\caption{Astrophysical scenarios for galaxy cluster accretion shocks assuming $\beta=100$ (except model D which assumes a constant magnetic field). Different columns show the scaling parameters of the magnetic 
field (columns $2,4,6$) and shock velocity (columns $3,5,7$) with cluster virial mass and redshift for four different models (see text). Scalings 
are of the form: $\mathcal{A}_i(M_{\rm vir}/10^{14}\,{\rm M}_{\odot})^{\alpha_i}\times \mathcal{E}(z)^{\beta_i}$, where $i=\{{\rm sh},{\rm mag}\}$ 
stands for the shock velocity and magnetic field cases, respectively, and $\mathcal{E}(z)=\sqrt{\Omega_{\rm M} \, (1+z)^3 + \Omega_\Lambda}$.}
\vspace{0.2cm}
\centering
\begin{tabular}{lcccccc}
\hline\hline
Model &  $\mathcal{A}_{\rm mag}$ & $\mathcal{A}_{\rm sh}$   & $\alpha_{\rm mag}$ & $\alpha_{\rm sh}$ & $\beta_{\rm mag}$ & $\beta_{\rm sh}$ \\
 & [$\mu$G] & [km\,s$^{-1}$] &  &  &  & \\
\hline
A & 0.145 & 838 & 0.31 & 1/3 & 4/3 & 0\\
B & 0.145 & 688 & 0.31 & 1/3 & 4/3 & 0\\
C & 0.068 & 1316 & 0.31 & 1/3 & 4/3 & 0\\
D & 1 & 1316 & 0 & 1/3 & 4/3 & 0\\
\hline\hline
\end{tabular}
\label{tab:models_vsh_B}
\end{table}
In the second scenario (model B), model A is modified replacing the infall shock velocity by that of Eq.~(\ref{eq:vsh_sim}), which has 
been motivated using cosmological simulations, leaving the magnetic field scaling unchanged. In the third scenario (model C), scalings are 
computed using both Eq.~(\ref{eq:vsh_sim}) for the shock velocity and $B_2$ for the magnetic field but, in this case, interpolating the simulated 
Mach number profile of Ref.~\cite{Rost24} at the mean distance from the cluster center, which corresponds to about $2.5\,R_{\rm vir}$. Finally, 
the fourth scenario (model D), postulates a constant magnetic field of $1\,\mu$G (i.e., independent of cluster mass) using the accretion shock 
velocity–mass scaling of model C. For all models, the magnetic field evolution with redshift follows from the pressure profile which, in turn, depends on cosmology. The scalings obtained for the four different models are presented in Table~\ref{tab:models_vsh_B}.

\subsection{\label{sec:MaxEn} Maximum energy}

Following Ref.~\cite{Kang:97} it is assumed that the CRs are accelerated at the shock fronts through the
first-order Fermi mechanism. Assuming that the diffusion coefficient of accretion shocks in galaxy clusters 
is proportional to the Bohm diffusion one, the following expression for the mean acceleration time is obtained \cite{Jokipii:87} 
\begin{equation}    
\tau_{\textrm{acc}}(E) = \frac{\chi}{\chi-1} \left( \frac{c}{v_{\textrm{sh}}}\right)^2 \frac{r_\textrm{g}(E)}{c}\, \eta \, 
\mathcal{J}(\theta,\chi,\eta),
\label{Lacc}
\end{equation}
where $r_\textrm{g}$ the gyroradius, $\theta$ is the angle between the magnetic field and the shock normal, $\eta$ is the 
proportionality constant between the diffusion coefficient parallel to the magnetic field and the Bohm diffusion coefficient, 
and  
\begin{equation}    
\mathcal{J}(\theta,\chi,\eta)=\cos^2\theta + \frac{\sin^2\theta}{1+\eta^2}+\chi \frac{\cos^2\theta + 
\chi^2 \, \frac{\sin^2\theta}{1+\eta^2}}{\left[ \cos^2\theta + \chi^2 \, \sin^2\theta \right]^{3/2} }.
%
\end{equation}
\begin{figure}[!ht]
\centering
\setlength{\abovecaptionskip}{0pt}
\includegraphics[width=7.8cm]{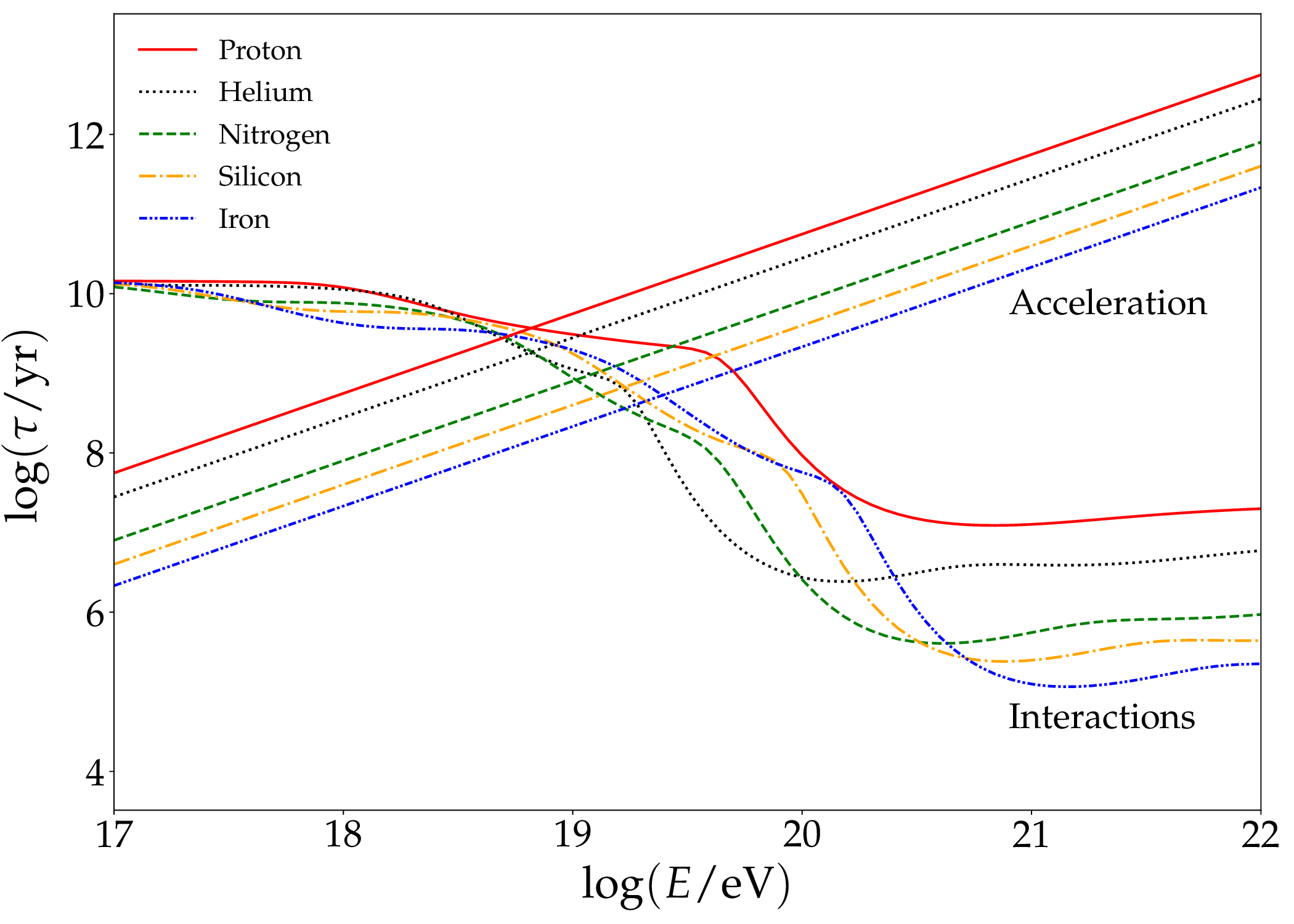}
\caption{Mean interaction and acceleration times as a function of the logarithm of primary energy for model C, $\eta=1$, 
$\theta=45^\circ$, and $\beta=100$ for all nuclear species considered.  \label{IntL}}
\end{figure}
A compression ratio $\chi=4$ is assumed, which corresponds to the case of a strong shock. Since, appropriate 
values of $\eta$ are in the $[1,10]$ interval \cite{Allard:09}, the values $\eta=1$ and $\eta=10$ are 
considered in the subsequent calculations. Recent findings indicate that, in addition to parallel shocks, 
CRs might also be efficiently accelerated in quasiperpendicular shocks \cite{Orusa:23}. Therefore, the values 
of $\theta$ considered in this work are: $0^\circ$, $45^\circ$, and $90^\circ$.

The CRs accelerated in galaxy cluster shocks lose energy mainly due to their interactions with 
the low-energy photons of the cosmic microwave background (CMB) and the extragalactic background light 
(EBL). Therefore, the maximum energy achieved by a CR accelerated in an accretion shock corresponding to 
a galaxy cluster of a given mass $M_{\rm vir}$ and redshift $z$ is estimated by solving the following 
equation 
\begin{equation}
\tau_{\textrm{int}}(E_\textrm{max},z,Z)=\tau_{\textrm{acc}}(E_\textrm{max},Z,M_{\rm vir}),
\end{equation}
\noindent where $\tau_{\textrm{int}}$ is the total mean interaction time and $Z$ is the charge number of the nuclear 
species under consideration. Figure \ref{IntL} shows the mean interaction time of the five nuclear species considered in 
this work: proton, helium, nitrogen, silicon, and iron at $z=0$ as a function of the logarithm of the energy, obtained 
from CRPropa \cite{CRPropa}. The mean interaction time shown in the figure includes the photo-pion production, photodisintegration, 
and pair production processes (see Appendix \ref{ApIntL} for details). The energy loss due to the adiabatic expansion 
of the Universe is also included but its effect on the maximum energy calculation is small. The EBL model used in the mean 
free path calculation is the one of Ref.~\cite{Gilmore:12}. The figure also shows the acceleration time obtained from 
Eq.~(\ref{Lacc}) for the five nuclear species considered in accretion model C and the quasiparallel case, $\theta=45^\circ$, 
for a cluster of $10^{15}\,$M$_\odot$. From the figure it can be seen that the maximum energy for the different nuclear types 
ranges from $\sim 10^{18.8}$ to $\sim 10^{19.4}$ eV. 
\begin{figure}[!ht]
\centering
\setlength{\abovecaptionskip}{0pt}
\includegraphics[width=7.8cm]{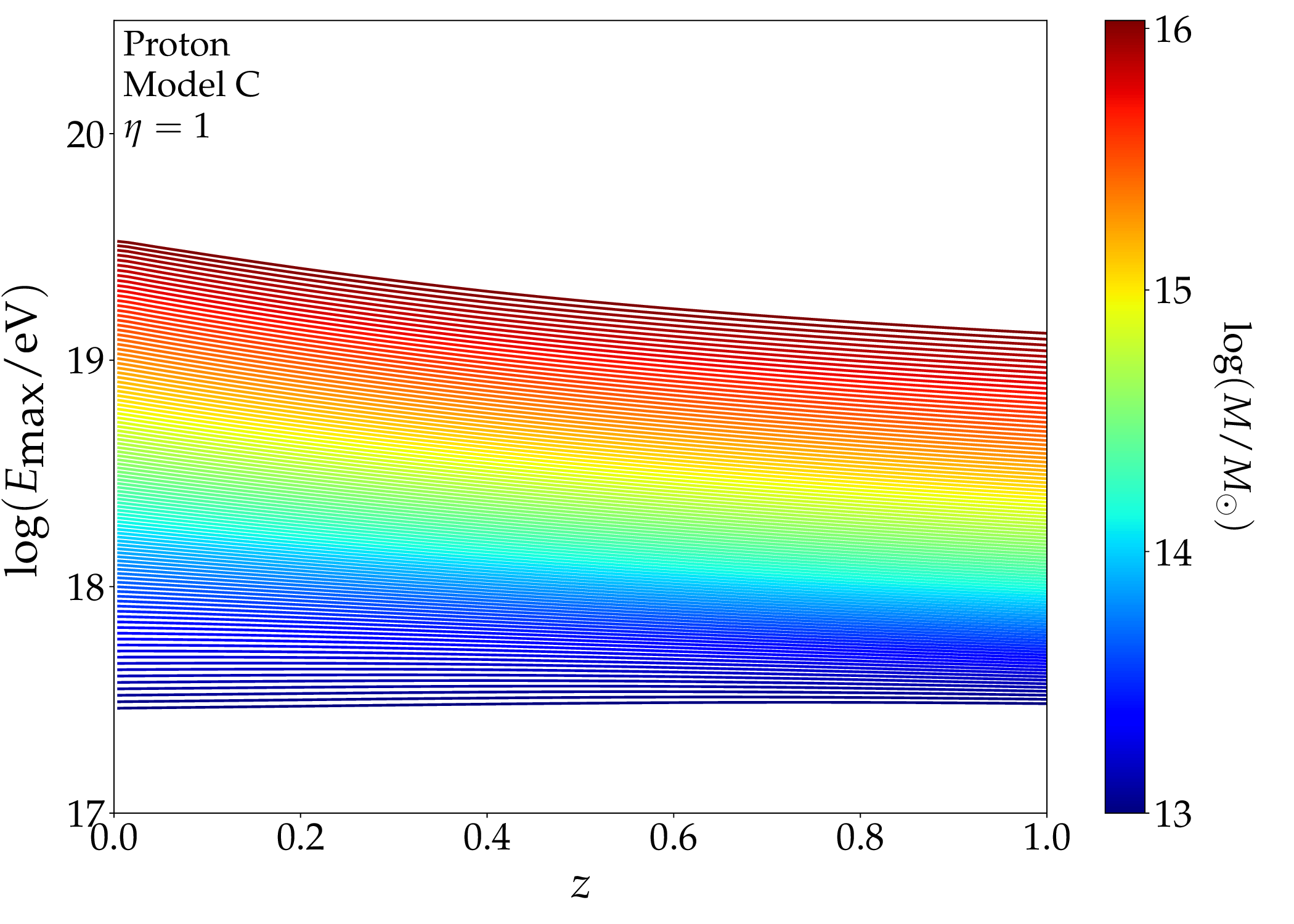}
\includegraphics[width=7.8cm]{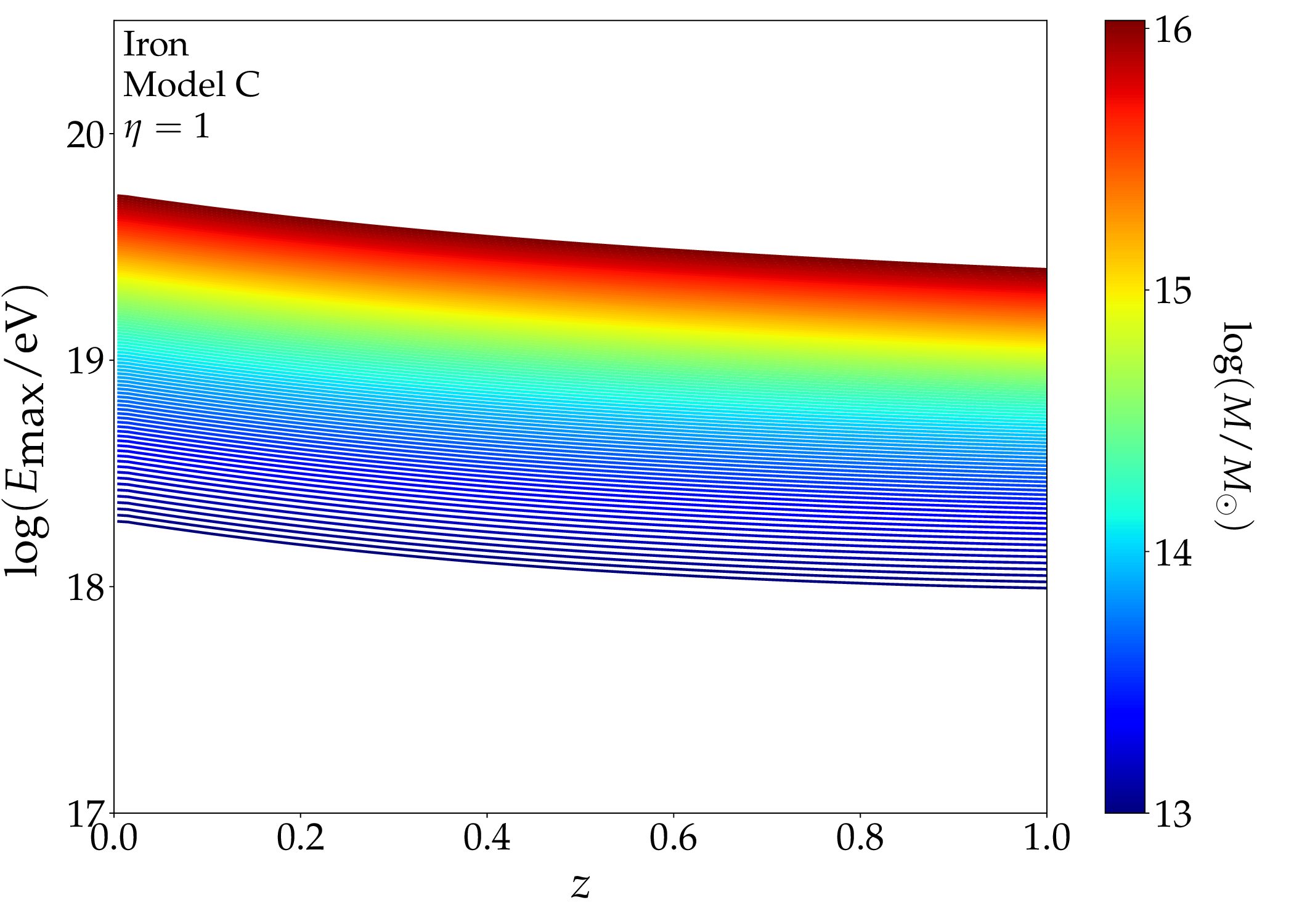}
\caption{Maximum energy as a function of the cluster redshift for different values of the cluster virial mass, 
for accretion shock model C, $\eta=1$, $\theta=45^\circ$, and $\beta = 100$ for protons (top panel), and iron 
nuclei (bottom panel). 
\label{MaxEnParC}}
\end{figure}
The maximum energy attained by the different CR types depends on the cluster virial mass and redshift. Figure \ref{MaxEnParC} 
shows the maximum energy corresponding to proton and iron primaries as a function of the redshift for different values of the 
cluster virial mass. The figure shows that the maximum energy varies more significantly with the cluster virial mass than with 
redshift. For any given redshift, the maximum energy can vary more than an order of magnitude for the different clusters, in 
contrast to the smaller variation observed at a fixed cluster mass during the redshift interval considered.

\subsection{Cosmic ray luminosity}

The differential number of CRs of mass number $A$ injected by the sources per unit of time and energy is modeled as a power law 
with a broken exponential cutoff \cite{AugerFit:23} 
\begin{equation}
\frac{dN_{\textrm{CR},A}}{dt dE}=C_0 \left( \frac{E}{E_0}\right)^{-\gamma} \left\lbrace 
\begin{array}{ll} 
1 & E<E_{\textrm{max}} \\[0.3cm] 
\exp\left( 1-E/E_{\textrm{max}} \right)  & E\geq E_{\textrm{max}} 
\end{array} 
\right. 
\label{InjSpec}
\end{equation}
where $C_0$ is a normalization constant, $\gamma$ is the spectral index, and $E_{\textrm{max}}\equiv E_{\textrm{max}}(z,M_{\rm vir},Z)$ 
is the maximum energy. Note that the spectral index is the same for all nuclear species considered. The normalization constant $C_0$ is 
determined from the CR luminosity which, in turn, is obtained from the accretion rate of clusters. In particular, the kinetic energy per 
unit time of the accretion shocks at a distance $r_{\rm sh}$ from the cluster center can be written as 
$\mathcal{L}=f_\textrm{b} G M_{\rm vir} \langle \dot{M} \rangle/r_\textrm{sh}$ \cite{Fang:16}, where 
$f_\textrm{b}=f_0\, (M_{\rm vir}/10^{15}\,{\rm M}_\odot)^{0.16}$ \cite{Gonzalez:13} is the average baryon fraction in clusters, 
$r_\textrm{sh}\gtrsim R_{\rm vir}\simeq 2.6 \times (M_{\rm vir}/10^{15}\,{\rm M}_\odot)^{1/3}$ Mpc \cite{Fang:16} is the shock radius 
and 
\begin{eqnarray}
\langle \dot{M} \rangle &\simeq& 1.01 \times 10^{5}\, \left( \frac{M_{\rm vir}}{10^{15}\,{\rm M}_\odot}\right)^{1.127} 
(1+1.17\, z) \nonumber \\
&&\mathcal{E}(z)\,{\rm M}_\odot \, \textrm{yr}^{-1},
\end{eqnarray}
\cite{McBride:09} is the mass accretion rate with the function $\mathcal{E}(z)=\sqrt{\Omega_\textrm{M} (1+z)^3+\Omega_\Lambda}$. 
In the model, the mass dependence of $f_0$ for the cosmology adopted in this work is also considered. In particular, $f_0=0.188$ is
used for clusters with $M_{500}>2\times10^{14}\,$M$_{\odot}$, whereas for less massive clusters with $M_{500}<2\times10^{14}\,$M$_{\odot}$ we take an average of $f_0=0.155$ around the observed scatter (see Ref.~\cite{Gonzalez:13}). 

Therefore, assuming that the CR luminosity is a fraction $f_\textrm{CR}$ of the accretion shock kinetic energy rate, the following 
expression is obtained
\begin{eqnarray}
\mathcal{L}_{\rm CR}\,(M_{\rm vir},z) &=&  \mathcal{L}_{46}
 \times 10^{46}\,\left( \frac{M_{\rm vir}}{10^{15} \, {\rm M}_\odot} \right)^{1.95} 
(1+1.17\, z) \nonumber \\
&&\mathcal{E}(z) \, f_\textrm{CR} \, \textrm{erg} \, \textrm{s}^{-1}{\rm ,}
\end{eqnarray}  

\noindent where $\mathcal{L}_{46}=1.98$ or $1.63$ if $f_0=0.188$ or $0.155$, respectively. The normalization constant in Eq.~(\ref{InjSpec}) 
is computed assuming that the luminosity corresponding to a nucleus of mass number $A$ is a constant fraction $I_A$ of the total CR 
luminosity, i.e.
\begin{equation}
\int_{E_{\textrm{min}}}^\infty dE \, \frac{dN_{\textrm{CR},A}}{dt dE} \, E = I_A \, \mathcal{L}_\textrm{CR}, 
\end{equation} 
where $E_{\textrm{min}}$ is the minimum energy of the CRs.

\subsection{The local galaxy cluster sample}

The galaxy cluster sample presented in Ref.~\cite{Hernandez24} is adopted to model the distribution of galaxy clusters in the local 
Universe. This work introduces a sample of the most massive galaxy clusters in our cosmic neighborhood together with a comprehensive, 
homogeneous list of cluster masses and distances from the Earth. The resulting selection is primarily based on systems from the Planck 
survey of Sunyaev-Zeldovich (SZ) galaxy clusters \cite{PlanckSZ14} which also have an entry in the Tully galaxy groups catalog 
\cite{Tully15}, with some additional clusters selected from various x-ray catalogs (see Ref.~\cite{Hernandez24} and references therein). 
In total, the cluster catalog comprises a collection of $45$ systems within a comoving distance of $227\,$Mpc from the observer and 
$M_{500}$ exceeding a mass cutoff of $M_{\rm low}\sim 2\times10^{14}\,$M$_{\odot}$. As noted in Ref.~\cite{Hernandez24}, although the 
selection process is guided by the availability of observational data, the combination of these catalogs should lead to a very complete 
coverage of massive local clusters. For the selection of local clusters, the SZ mass estimate provided by Ref.~\cite{Hernandez24} is used 
whenever available. In cases where the latter is not accessible, the x-ray measurement is used. If neither SZ nor x-ray estimates are 
available, a corrected dynamical mass estimate obtained from the Tully galaxy groups catalog is used. Subsequently, all available $M_{500}$ 
cluster masses are converted 
to $M_{\rm vir}$ by assuming an NFW profile and the halo concentrations of Ref.~\cite{Duffy08}. Using this procedure the condition 
$M_{500}\gtrsim M_{\rm low}$ translates into the constraint $M_{\rm vir}\gtrsim M_{\rm vir,low}$, with $M_{\rm vir,low}=10^{14.57}\,$M$_{\odot}$, 
which, within uncertainties, is satisfied for about $70\%$ of the systems within $150\,$Mpc. Additionally, the observed cluster distances 
reported in Ref.~\cite{Hernandez24} are considered, except in the cases of Virgo and Fornax where projection effects in the sky introduce 
systematic errors leading to unreliable distance determinations. Instead, for Virgo, the value given in Ref.~\cite{Mei07} is adopted, which 
conducts a detailed analysis of the distance of about a hundred early type galaxies in the cluster, resulting in a better distance measurement. 
In the case of Fornax, the recent tip of the giant branch distance measurement provided by Ref.~\cite{Anand24} is considered. The list of 
all massive clusters within $150\,$Mpc provided by~\cite{Hernandez24} is shown in Table~\ref{tab:local_clusters}.
\begin{table}[!th]
\caption{Local sample of massive galaxy clusters within $150\,$Mpc from Earth sorted by increasing distance according to Ref.~\cite{Hernandez24}. 
From left to right columns show cluster name/ID, comoving distance, our virial mass estimate (see text) and the observational method used for 
$M_{500}$ determination. 
}
\vspace{0.2cm}
\centering
\begin{tabular}{lccc}
\hline\hline
Name & $d$ & $M_{\rm vir}$ & Method\\
& [Mpc] & [$10^{14}\,$M$_{\odot}$] \\
\hline
Virgo           &  $16.70 \pm 0.20^{\dagger}$   &  $9.16 \pm 1.10$ & SZ\\ 
Fornax/AS0373   &  $19.30 \pm 0.70^{\dagger}$   &  $1.77_{-0.77}^{+0.62}$   & X\\ 
Centaurus/A3526 &  $56.64 \pm 27.40$   &  $2.26 \pm 0.21$          & SZ\\ 
Hydra/A1060     &  $63.30 \pm 27.36$   &  $3.49_{-1.45}^{+0.89}$   & X\\ 
AWM7            &  $70.31 \pm 27.31$   &  $7.04_{-3.30}^{+1.80}$   & X\\ 
Norma/A3627     &  $76.02 \pm 25.35$   &  $4.80  \pm 0.55$         & SZ\\ 
Perseus/A426    &  $78.31 \pm 27.31$   &  $8.76_{-4.21}^{+2.32}$   & X\\ 
A347            &  $82.36 \pm 27.33$   &  $6.05$                   & dyn\\
3C129           &  $91.37 \pm 26.75$   &  $7.73_{-3.43}^{+3.37}$   & X\\ 
A3581           &  $103.50 \pm 25.27$  &  $3.41  \pm 0.08$         & SZ\\ 
A1367           &  $103.78 \pm 25.27$  &  $3.27  \pm 0.27$         & SZ\\ 
A2877           &  $105.26 \pm 27.22$  &  $2.05  \pm 0.26$         & SZ\\ 
Coma/A1656      &  $109.30 \pm 20.92$  &  $10.22 \pm 0.40$         & SZ\\ 
A539            &  $129.97 \pm 25.26$  &  $3.75_{-1.23}^{+0.47}$   & X\\ 
A2634           &  $137.52 \pm 27.15$  &  $6.43_{-1.47}^{+0.99}$   & X\\ 
A2197/99        &  $140.23 \pm 25.17$  &  $5.25 \pm 0.25$          & SZ\\ 
A496            &  $150.06 \pm 25.16$  &  $5.17 \pm 0.33$          & SZ\\ 
\hline\hline
$^{\dagger}$Refs.~\cite{Mei07} and \cite{Anand24}.
\end{tabular}
\label{tab:local_clusters}
\end{table}

To approximately reproduce the expected number of clusters in our vicinity, we select all clusters with masses above $M_{\rm low}$ within a comoving distance of $110\,$Mpc, which for the cosmology adopted in this work corresponds to a redshift of $z_0=0.02485$. The resulting sample comprises a total of $8$ objects that is consistent at the $\sim 1\,\sigma$ level (assuming Poissonian uncertainties) with the expected number of massive clusters at $z \leq z_0$ determined by the halo mass function (HMF) of Ref.~\cite{Tinker08} after a proper scaling to the cosmology adopted in this paper. As a result, the final selection includes some of the most iconic structures present in our cosmic neighborhood such as the Virgo, Hydra, Perseus and Coma galaxy clusters, among others. 

\begin{figure}[t]
\centering
\setlength{\abovecaptionskip}{0pt}
\includegraphics[width=7.5cm]{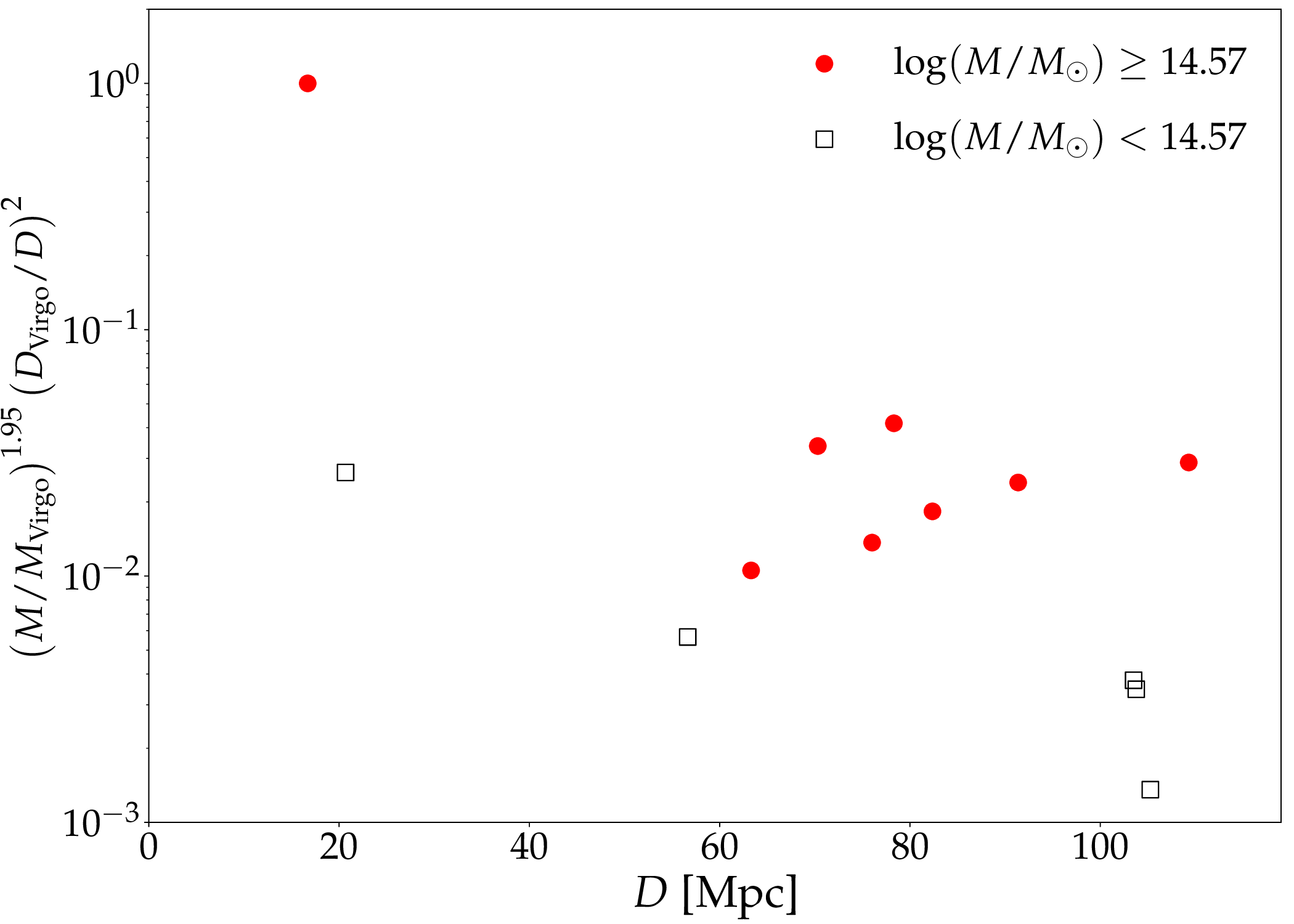}
\caption{The weight of each galaxy cluster in the local sample as a function of comoving distance. The masses and distances 
are normalized to the corresponding values of Virgo. The galaxy clusters are at comoving distances smaller than $110\,$Mpc.
\label{ClWeight}}
\end{figure}

It is generally believed that Virgo may account for a significant fraction of the UHECRs arriving at Earth due to its relative proximity (see, e.g., Ref.~\cite{Ptuskin:19}), compared to other massive but more distant clusters like Perseus or Coma. However, a proper gauging of the relative contribution of the different clusters to the measured CR flux on Earth has to be performed considering cluster masses and distances in a consistent way. Figure~\ref{ClWeight} shows the weighted contribution of individual clusters in Table~\ref{tab:local_clusters} normalized to the corresponding Virgo values as a function of distance. Each cluster is weighted taking into account the mass dependence of the CR luminosity divided by the distance of the cluster squared. This simple calculation indicates that, within this scenario, the Virgo cluster is the dominant CR source, with its weight exceeding that of the rest of the sample by more than an order of magnitude. However, a more rigorous calculation that accounts for both propagation of the different nuclear species and cosmology is ultimately needed, and it will be presented below. 

It is worth mentioning that the discrete local cluster sample is approximately complete only for the high-mass end of the cluster mass distribution within a comoving distance of about $110\,$Mpc. To take into account the existence of clusters below the mass cutoff imposed in Ref.~\cite{Hernandez24} that may contribute to the overall UHECR flux, the universal, redshift-dependent HMF of Ref.~\cite{Tinker08} is used to account for clusters with virial masses below the cutoff down to a minimum mass, typical of galaxy groups. In this way, the possibility that galaxy groups in the local Universe also develop an envelope of less-energetic accretion shocks where CRs can accelerate is also considered. As a result, local sources in the model contains a discrete contribution corresponding to well-know, more massive galaxy clusters located in our vicinity, plus a statistical contribution of less-massive systems modeled following the abundance dictated by the HMF evolution in the redshift range $z=[0,z_0]$. Further details of this procedure will be provided in the next section.

\section{\label{sec:Charac} Characterization of Cosmic Ray flux and composition}

\subsection{\label{sec:FluxCompo} Flux and composition profile calculation}

The propagation of the CRs is affected by their interactions with the low-energy photons of 
the radiation field that fills the intergalactic medium. The most important photon backgrounds are 
the CMB and the EBL. The main processes undergone by these particles during propagation are: 
electron-positron pair photoproduction, pion photoproduction, and photodisintegration of nuclei. 
Considering a uniform distribution of galaxy clusters, the flux at Earth corresponding to nuclei of 
mass number $A$ for galaxy clusters located at redshifts between $z_{\textrm{a}}$ and $z_{\textrm{b}}$ 
with virial masses between $M_\textrm{a}$ and $M_\textrm{b}$ (for simplicity, the subscript ``vir'' will 
be omitted from the masses in what follows) is given by
\begin{widetext}
\begin{equation}
J_{\textrm{U},A}(E,z_{\textrm{a}},z_{\textrm{b}},M_\textrm{a},M_\textrm{b}) = 
\frac{c}{4\pi} \sum_{A'} \int_{z_{\textrm{a}}}^{z_{\textrm{b}}} \! \! \!  dz' 
\int_{E_{\textrm{min}}}^\infty \! \! \! dE' \int_{M_\textrm{a}}^{M_\textrm{b}} \! \! \!  dM \
\frac{Q_{A'}(E',M,z')}{(1+z')\, H(z')}\, \frac{d\bar{n}_{AA'}}{dE}(E;E',z'),  
\label{JAcont}
\end{equation}
\end{widetext}
where $c$ is the speed of light and $H(z)=H_0 \, \mathcal{E}(z)$ is the Hubble parameter with its value at $z=0$ being $H_0=67.3$ 
km~s$^{-1}$~Mpc$^{-1}$. Here $d\bar{n}_{AA'}(E;E',z')/dE$ is the differential fraction of particles reaching the Earth with energy 
$E$ and mass number $A$ generated by a nucleus emitted by a galaxy cluster at a redshift $z'$ with energy $E'$ and mass number $A'$. 
Also $Q_{A'}$ in Eq.~(\ref{JAcont}) is the number of nuclei with mass number $A'$ injected by the clusters per unit of energy, 
volume and time, which can be written as
\begin{equation}
Q_{A'}(E',M,z')=\frac{dN_{\textrm{CR},A'}}{dt' dE'}(E',z',M) \frac{dn_\textrm{CL}}{dM}(M;z'),
\end{equation} 
where $dn_\textrm{CL}/dM$ is the HMF given by Ref.~\cite{Tinker08}.

The distribution $d\bar{n}_{AA'}/dE$ is calculated from Monte Carlo simulations performed by using the CRPropa program \cite{CRPropa}. Also, in this case, the EBL model used in the simulations is the one of Ref.~\cite{Gilmore:12}. The simulations are performed considering the five different nuclear species mentioned before. CRs are injected in the intergalactic medium at redshift values obtained by sampling a uniform distribution in light-travel distance. Since the fit of the CR data is performed above $10^{18.7}$ eV redshift values in the interval $[0,1]$ are considered \cite{AugerFit:17}. The energy of the injected particles is quantized in discrete steps, each with a width of $\Delta \log(E/\textrm{eV})=0.01$. It ranges from $10^{18}$ to $Z'\times10^{21.5}$ eV, where $Z'$ is the charge number of the injected nuclei.

The flux at Earth from nuclei with mass number $A$ for a discrete set of nearby massive clusters is given by
\begin{widetext}
\begin{equation} 
J_{\textrm{D},A}(E) = \sum_i \sum_{A'} \frac{1}{(4 \pi)^2 
\, r_i^2 \, (1+z_i)} \int_{E_{\textrm{min}}}^\infty  
dE' \, \frac{dN_{\textrm{CR},A'}}{dt' dE'}(E',z_i,M_i) \frac{d\bar{n}_{AA'}}{dE}(E;E',z_i),
\end{equation}
\end{widetext}
where $r_i$, $z_i$, and $M_i$ are the comoving distance, the redshift, and the virial mass of the $i$th cluster of the catalog, respectively. In this case, a set of dedicated simulations are performed in order to calculate $d\bar{n}_{AA'}(E;E',z_i)/dE$ with more statistics. The only difference between these simulations and the previous ones is that, in the latter case, CRs are injected at the positions of each cluster listed in the catalog of Table~\ref{tab:local_clusters}.   

Since the catalog is obtained for cluster virial masses larger than about $M_\textrm{vir,low}$, the total flux at Earth 
corresponding to a nucleus of mass number $A$ is given by
\begin{eqnarray}
J_A(E) &=& J_{\textrm{D},A}(E) + 
J_{\textrm{U},A}(E,z_\textrm{min},z_\textrm{max},M_\textrm{min},M_\textrm{vir,low})  \nonumber \\  
&& + \, J_{\textrm{U},A}(E,z_0,z_\textrm{max},M_\textrm{vir,low},M_\textrm{max}), 
\end{eqnarray}

\noindent where $z_\textrm{min}=0$ and $z_\textrm{max}=1$ are the minimum and maximum redshifts, respectively. For 
the cluster virial mass range it is assumed that accretion shocks are present from the less-massive galaxy groups 
to the largest clusters. For $M_\textrm{min}$ two values are considered: $10^{13}$ and $10^{13.5}\,{\rm M}_\odot$, 
which are typical galaxy group mass scales, while $M_\textrm{max}=10^{16}\,{\rm M}_\odot$ is taken as an upper limit 
for the integration. However, due to the HMF distribution, the more massive clusters that contribute significantly to 
the flux are always less massive than $M_\textrm{max}$. 

The data to be fitted includes the total flux at Earth, the mean value of the natural logarithm of the mass number 
and its variance, which are given by
\begin{eqnarray}
J(E)&=& \sum_A J_A(E), \\
\langle \ln A \rangle(E) &=& \frac{\sum_A J_A(E) \ln A}{J(E)}, \\
\textrm{Var}[\ln A ](E) &=& \frac{\sum_A J_A(E) \ln^2 A}{J(E)} - \langle \ln A \rangle(E)^2.
\end{eqnarray}

\subsection{Fit to experimental data}

As mentioned before, the flux and the two first moments of $\ln A$ are fitted considering five nuclear
species injected by the sources. The free fitting parameters, $\gamma$, $f_{\textrm{CR}}$ and the five 
luminosity fractions $I_A$ (which are not independent since $\sum_A I_A = 1$) are estimated minimizing 
a likelihood function, which is composed by three $\chi^2$ terms for the flux and two first moments of
$\ln A$ and a Poisson term with $n=0$ for the two upper limits on the flux reported by Auger at the two 
highest energy bins considered. The flux data is taken from Ref.~\cite{AugerFlux:21}, the upper limits 
on the flux are taken from Ref.~\cite{AugerFluxUL:20}, and the first two moments of $\ln A$ data are 
taken from Ref.~\cite{AugerXmax:19}. Since simulated atmospheric showers are required to obtain the first 
two moments of $\ln A$ from the $X_\textrm{max}$ parameter, the data generated using the high-energy 
hadronic interaction models EPOS-LHC \cite{EPOSLHC:15} and Sibyll 2.3c \cite{Sibyll2.3c:19} are considered 
in this analysis. As mentioned before, the Auger data is fitted above $10^{18.7}$ eV. Since this study 
focuses on the potential of accretion shocks in galaxy clusters to explain the most energetic CRs, the 
origin of the ankle lies beyond the scope of this paper. It is important to remark that the Auger 
directional exposure is not considered in the calculation of the contribution from discrete sources. 
Instead, it is assumed that the flux measured by Auger accurately reflects the flux that would be 
observed by an observatory with uniform exposure across the entire sky. While this assumption introduces 
a limitation to the analysis, the primary goal of this study is to explore the connection between UHECRs 
and galaxy cluster accretion shocks, rather than to develop a detailed model of the Auger flux and 
composition measurements.
\begin{figure}[!ht]
\centering
\setlength{\abovecaptionskip}{0pt}
\includegraphics[width=8cm]{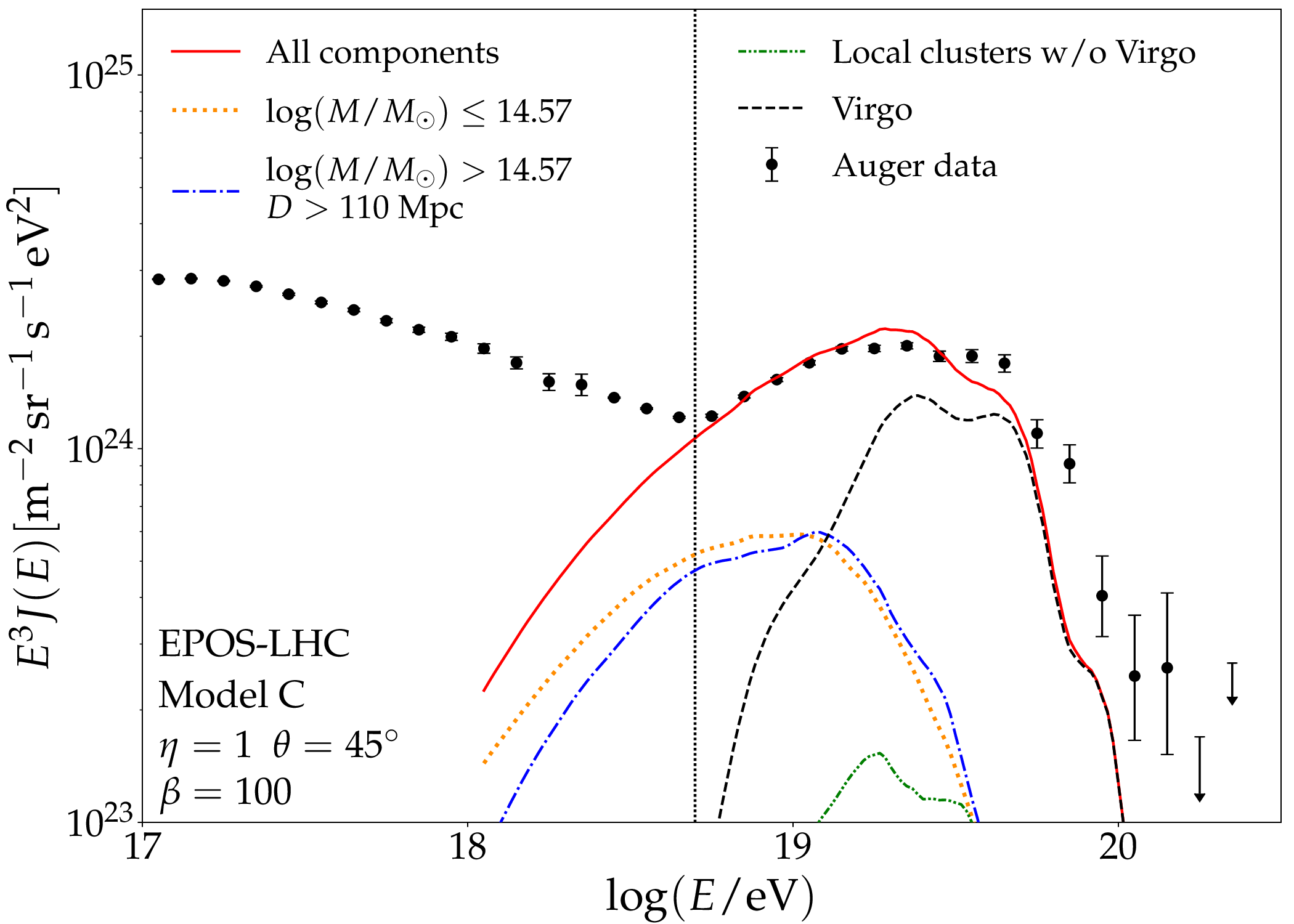}
\includegraphics[width=8cm]{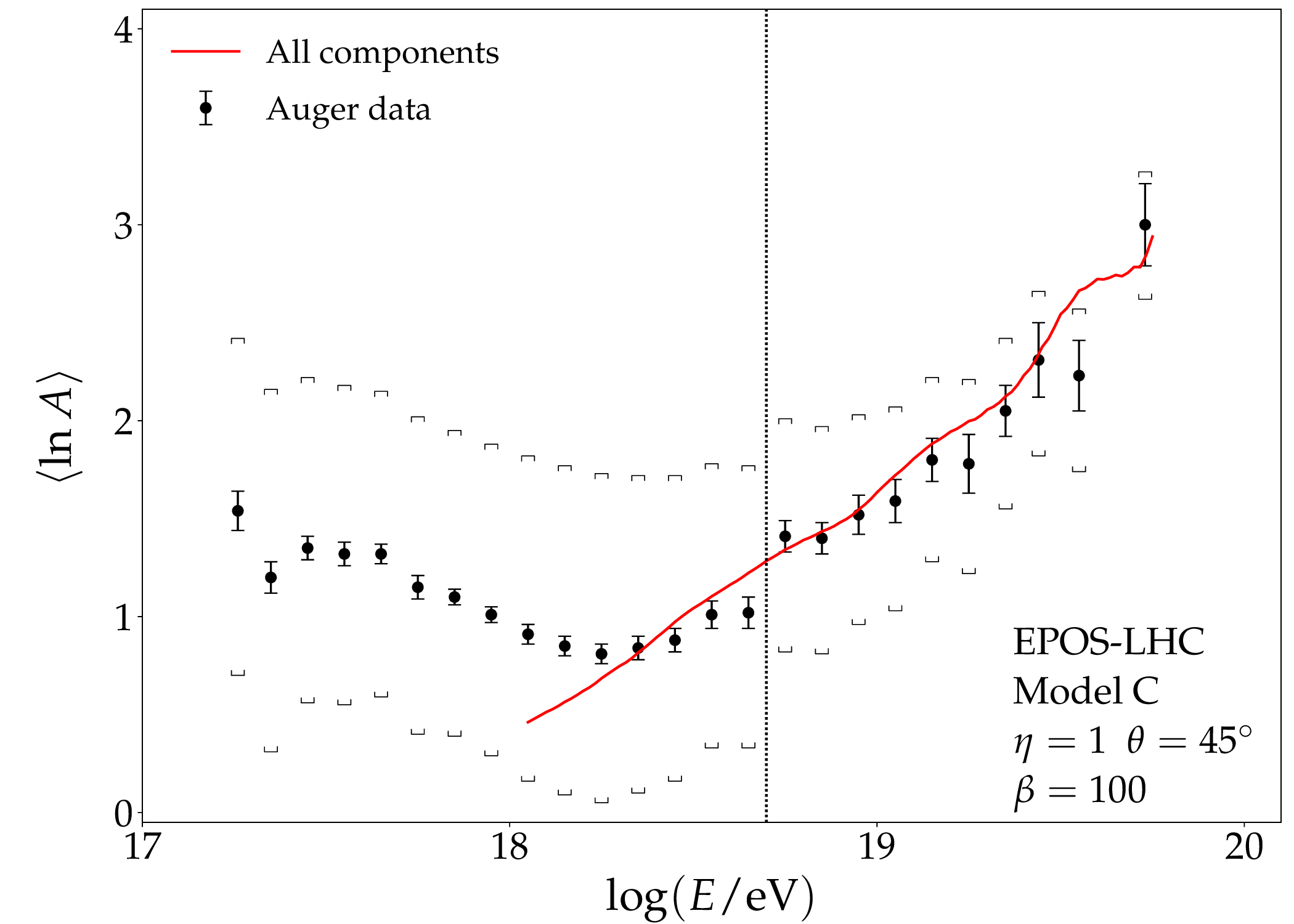}
\includegraphics[width=8cm]{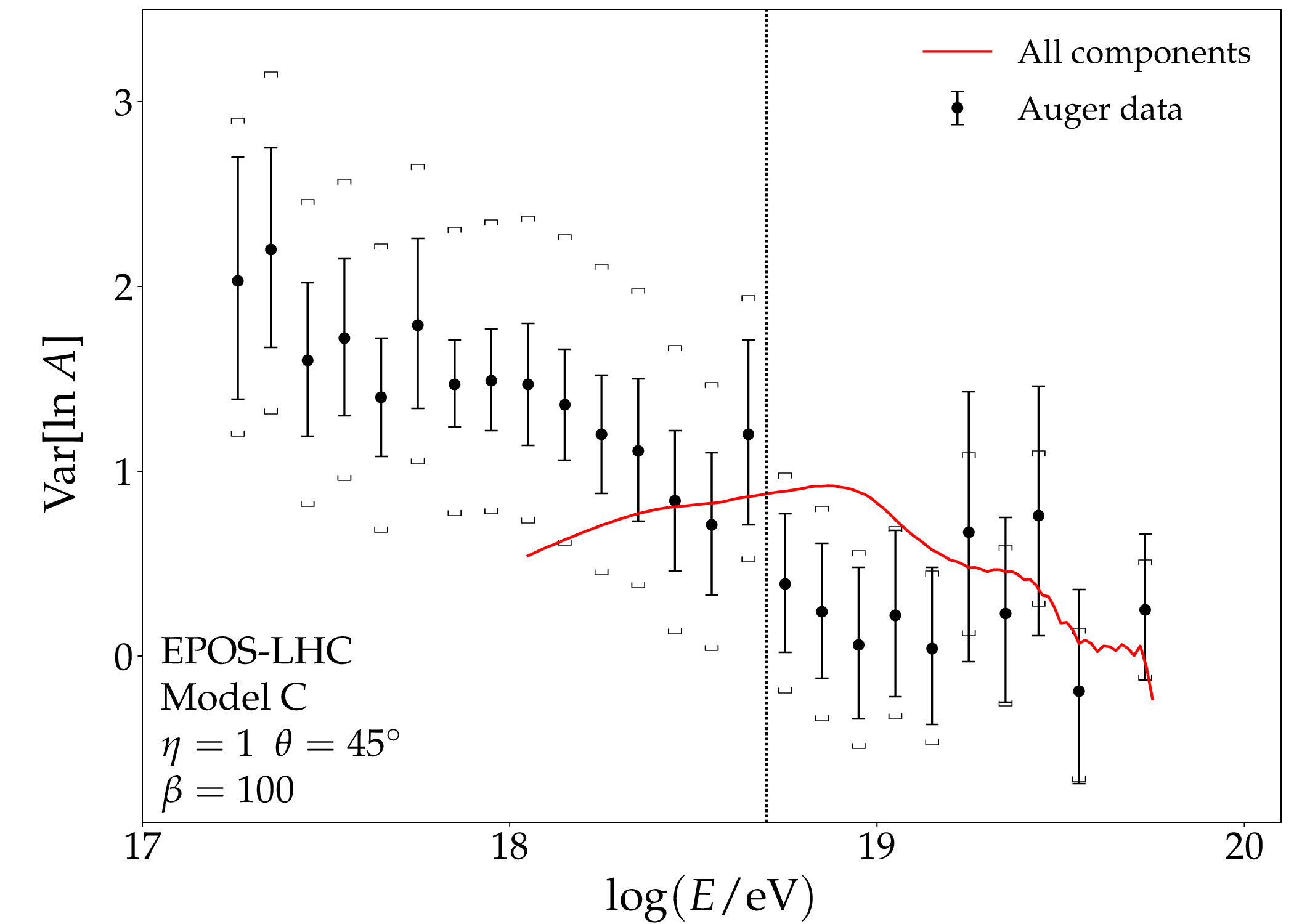}
\caption{Top panel: the cosmic ray flux, multiplied by the cube of the energy, as a function of the logarithm of the 
primary energy. Different cluster contributions to the total flux are shown (see text for details). Middle panel: mean 
value of $\ln A$ as a function of the logarithm of primary energy. Bottom panel: variance of $\ln A$ as a function of 
the logarithm of primary energy. In the three plots the data points represent the Auger measurements and the red solid 
line our best-fit model for $\log(M_\textrm{min}/{\rm M}_\odot)=13$ and $\beta=100$. The vertical lines mark the lower 
energy limit of the data used in the fit. The brackets in the composition data represent the systematic uncertainties.
\label{QParalellB100}}
\end{figure}
%
%
Figure \ref{QParalellB100} shows the fit of our model to the Auger data for quasiparallel shocks 
($\theta=45^\circ$) in the case of accretion shock model C and high-energy hadronic interaction model 
EPOS-LHC for $\eta=1$, $\beta = 100$ and $\log(M_\textrm{min}/{\rm M}_\odot)=13$. As anticipated, the 
highest energy portion of the flux is dominated by the contribution of the Virgo cluster, with other 
nearby galaxy clusters playing a comparatively minor role. Contributions from local Universe clusters 
with virial masses below $10^{14.57}\,$M$_\odot$, and those with masses above $10^{14.57}\,$M$_\odot$ 
but at redshifts $z=[z_0,1]$, are only significant well below the flux suppression energy. Crucially, 
this figure reveals that this model struggles to describe the flux suppression region, which is attributable 
to the low maximum energies obtained in the scenario under consideration. Conversely, the model successfully 
fits the composition data. The variance of $\ln A$ between $10^{18.7}$ and $10^{19.2}\,$eV is not perfectly 
reproduced, but the data remain consistent with the model within the total (statistical and systematic) 
uncertainties. Table \ref{tab:FitPB100} shows the best fit parameters for the case shown in 
Fig.~\ref{QParalellB100}.
\begin{table}[t!h]
\caption{Best-fit parameters for the model of Fig.~\ref{QParalellB100}.}.
\vspace{0.2cm}
\centering
\begin{tabular}{lcl}
\hline\hline
$\gamma$        & = & $-1.33 \pm 0.05$ \\
$f_\textrm{CR}$ & = & $(1.099 \pm 0.004)\times 10^{-2}$ \\
$I_\textrm{p}$  & < & $2.7 \times 10^{-6}$ \\
$I_\textrm{He}$ & = & $0.54 \pm 0.01$ \\
$I_\textrm{N}$  & = & $0.41 \pm 0.01$ \\
$I_\textrm{Si}$ & = & $0.06 \pm 0.02$ \\
$I_\textrm{Fe}$ & < &$0.0069$ \\
\hline\hline
\end{tabular}
\label{tab:FitPB100} 
\end{table}

\begin{figure}[t]
\centering
\setlength{\abovecaptionskip}{0pt}
\includegraphics[width=7.8cm]{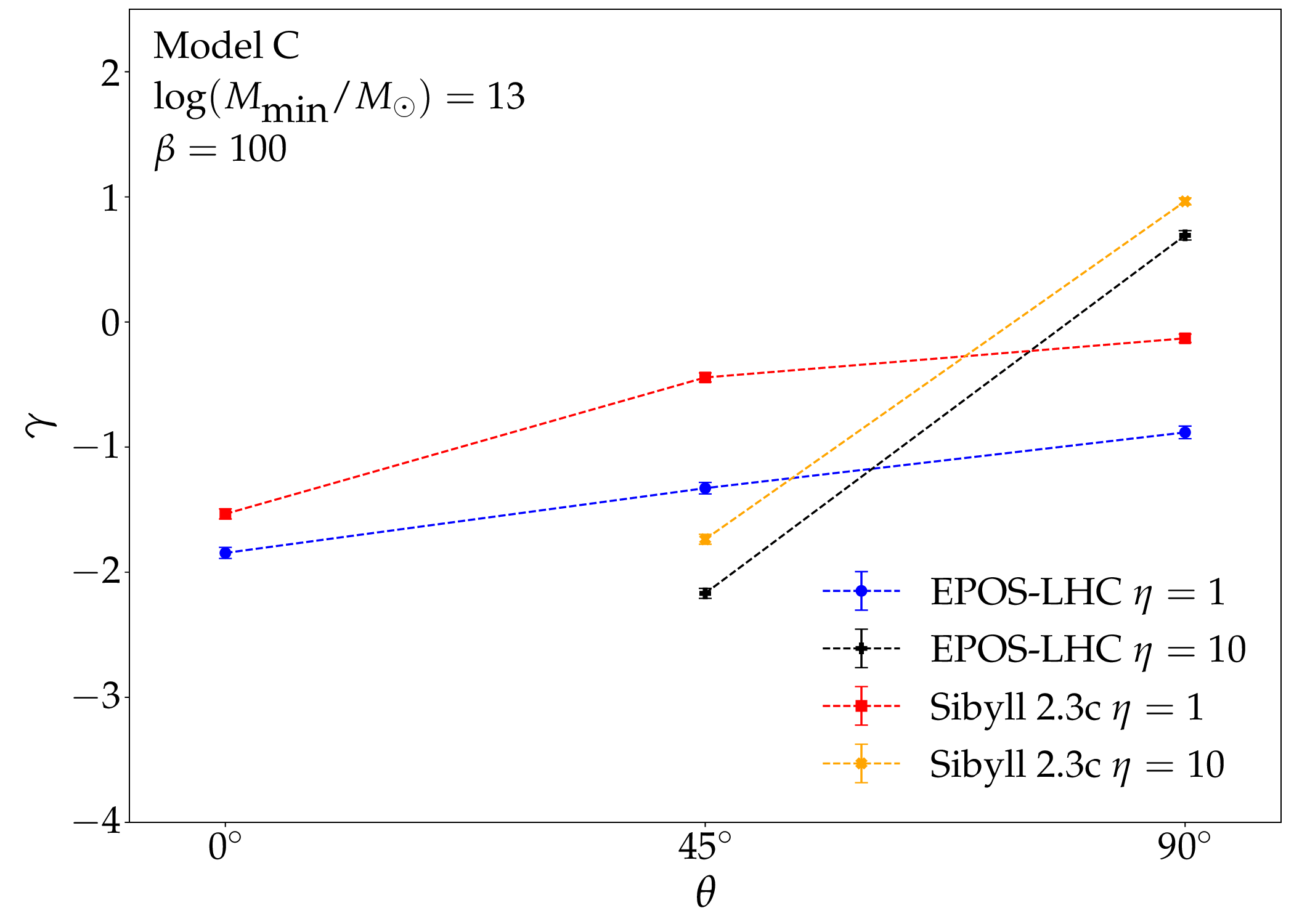}
\includegraphics[width=7.8cm]{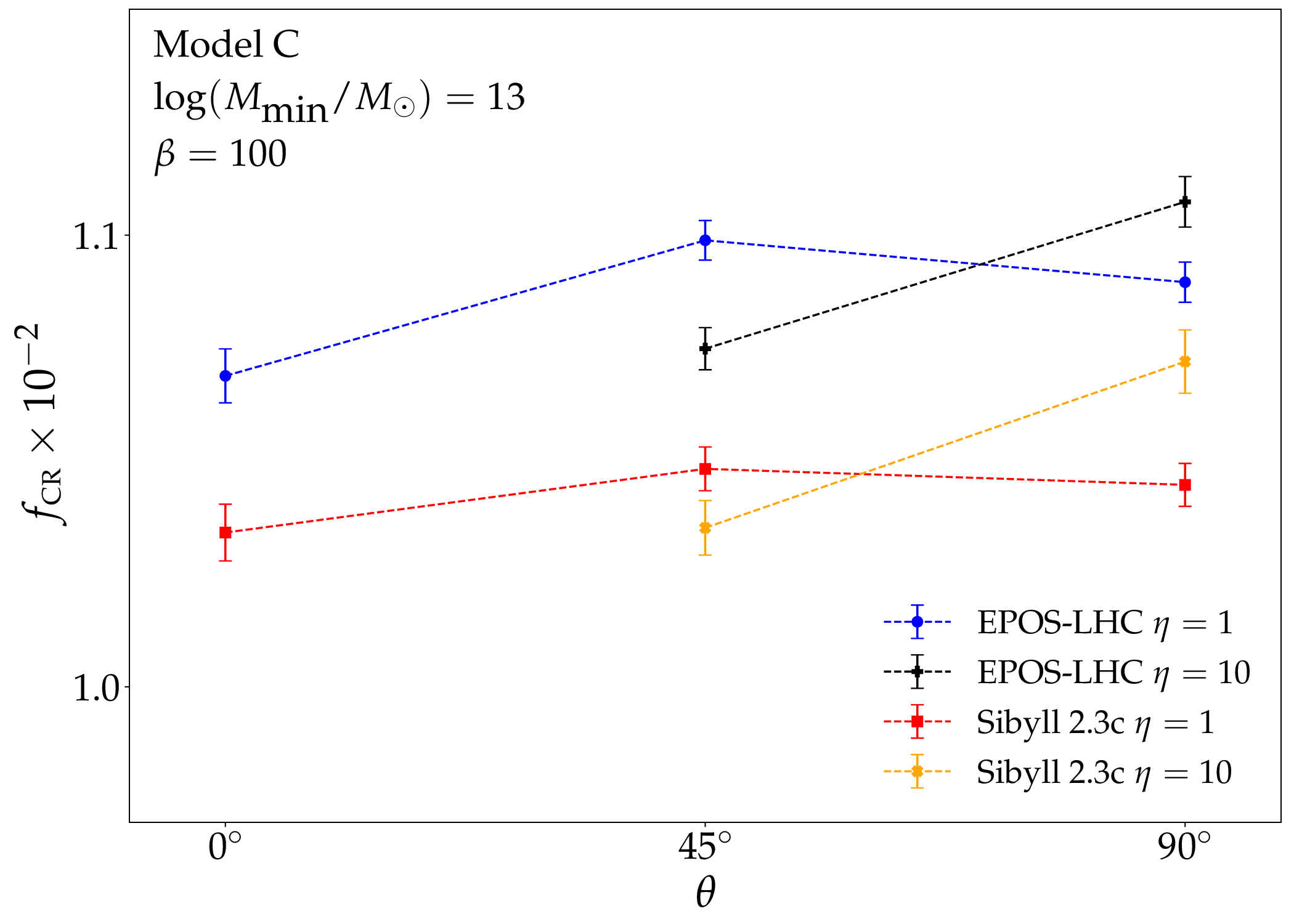}
\caption{Spectral index (top panel) and cosmic ray fraction (bottom panel) as a function of $\theta$ for all cases under 
consideration, except $\theta=0^{\circ}$ and $\eta=10$. The minimum value of the galaxy cluster mass considered here is 
$\log(M_\textrm{min}/{\rm M}_\odot)=13$ and $\beta=100$.
\label{ParFitB100}}
\end{figure}

The top panel of Fig.~\ref{ParFitB100} presents the best-fit spectral index of the injected spectrum as a function of 
the angle $\theta$ for accretion shock model C with $\log(M_\textrm{min}/{\rm M}_\odot)=13$. The results are shown for 
all considered angles and high-energy hadronic interaction models, except for the case of parallel shocks with $\eta = 10$, 
for which the maximum attainable energies are too low to perform a fit. The figure shows that $\gamma$ increases with 
the angle between the magnetic field and the shock normal, generally ranging between $-2.2$ and $1$. This trend arises 
because the maximum energy increases with $\theta$, thus requiring larger values of $\gamma$ to reach higher energies. 
The resulting $\gamma$ values further underscore the difficulty in reaching the suppression region of the flux. Notably, 
$\gamma$ values obtained for Sibyll 2.3c are larger than those for EPOS-LHC. This difference occurs because Sibyll 2.3c 
infers a heavier composition, and since heavier primaries attain larger maximum energies, less negative $\gamma$ values 
are required for this high-energy hadronic interaction model. The bottom panel of Fig.~\ref{ParFitB100} presents the 
best-fit values of the CR fraction parameter, $f_\textrm{CR}$, as a function of $\theta$. The plot reveals $f_\textrm{CR}$ 
values of about $0.01-0.011$ across different angles. Interestingly, despite the inability of the model to account for 
the suppression region, these $f_\textrm{CR}$ values align with the results obtained in Ref.~\cite{Inoue:07}. Note that, 
since these models do not reproduce the flux in the suppression region, the corresponding best-fit parameters do not 
yield an unbiased estimate of the underlying acceleration spectrum. In particular, the fits tend to favor smaller 
$\gamma$ values, including negative ones, to compensate for the low maximum energies, thereby biasing the inferred 
spectral index if UHECRs are indeed accelerated in these types of sources.

The maximum-energy values obtained for models A, B, and C are broadly similar across all primary species, following the 
ordering $E_{\textrm{max,C}} > E_{\textrm{max,A}} > E_{\textrm{max,B}}$. As a result, only model C and model A with 
$\eta = 10$ and $\theta = 90^\circ$ for Sibyll2.3c achieve a rough consistency with the suppression region (see 
Appendix~\ref{ApModelCSib} for model C). In all other scenarios considered, the maximum energies are too low to accurately 
reproduce the flux suppression. Although the variance of the composition is not perfectly matched in most cases, the overall 
composition trends remain compatible with the data within total uncertainties. It is important to note that the efficiency 
of perpendicular shocks in accelerating ions remains an open question (see, however, Ref.~\cite{Orusa:23}). For the models 
that roughly fit the suppression region, the fitted spectral indexes are $0.96 \pm 0.03$ for model C and $0.90 \pm 0.03$ 
for model A, both corresponding to $\theta = 90^\circ$, $\eta = 10$, and Sibyll2.3c. These results suggest that the inferred 
values of $\gamma$ are influenced by the limited maximum energies, the larger spectral index obtained for model C is consistent 
with its higher maximum energy relative to model A.

\begin{figure}[!ht]
\centering
\setlength{\abovecaptionskip}{0pt}
\includegraphics[width=8cm]{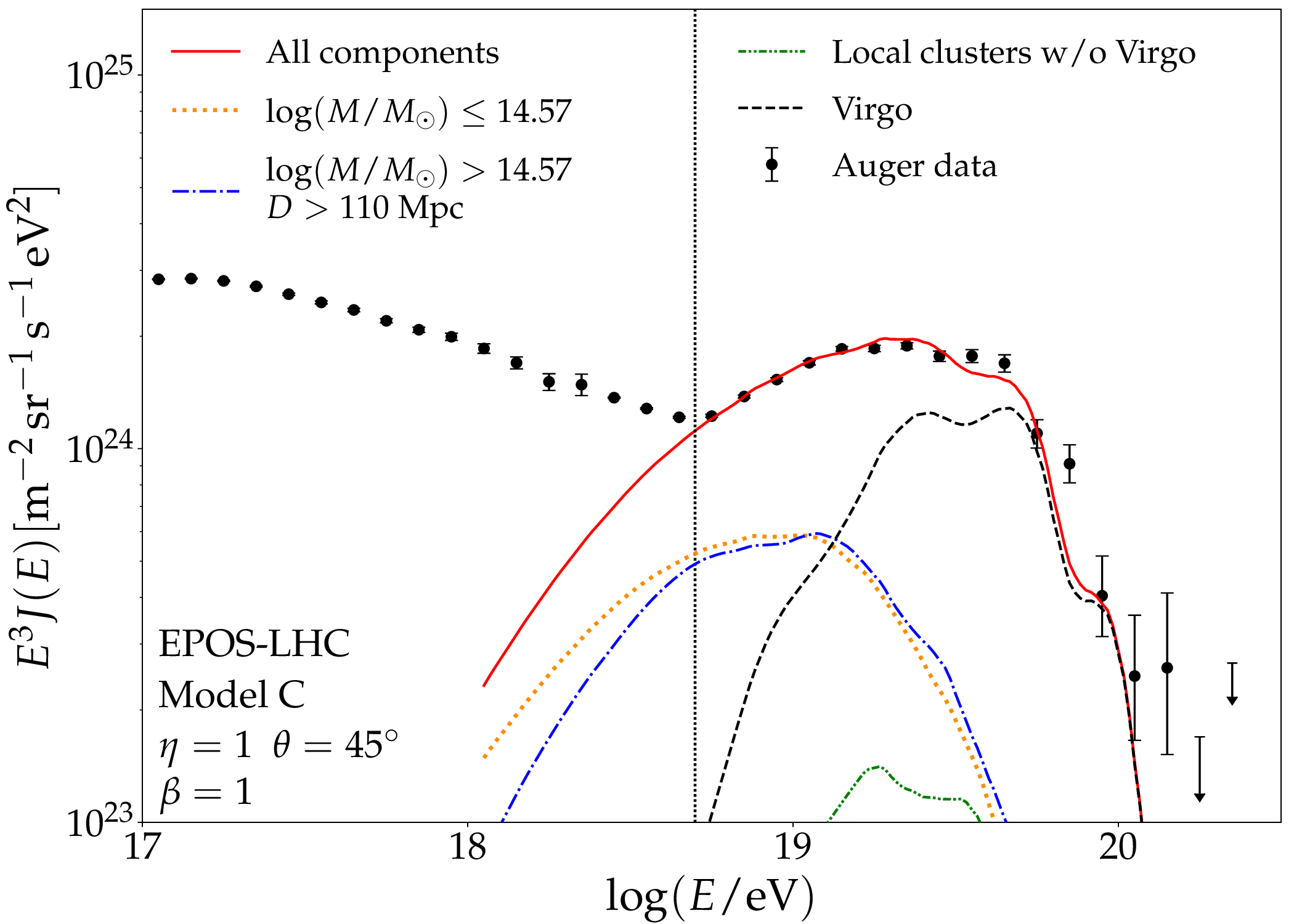}
\includegraphics[width=8cm]{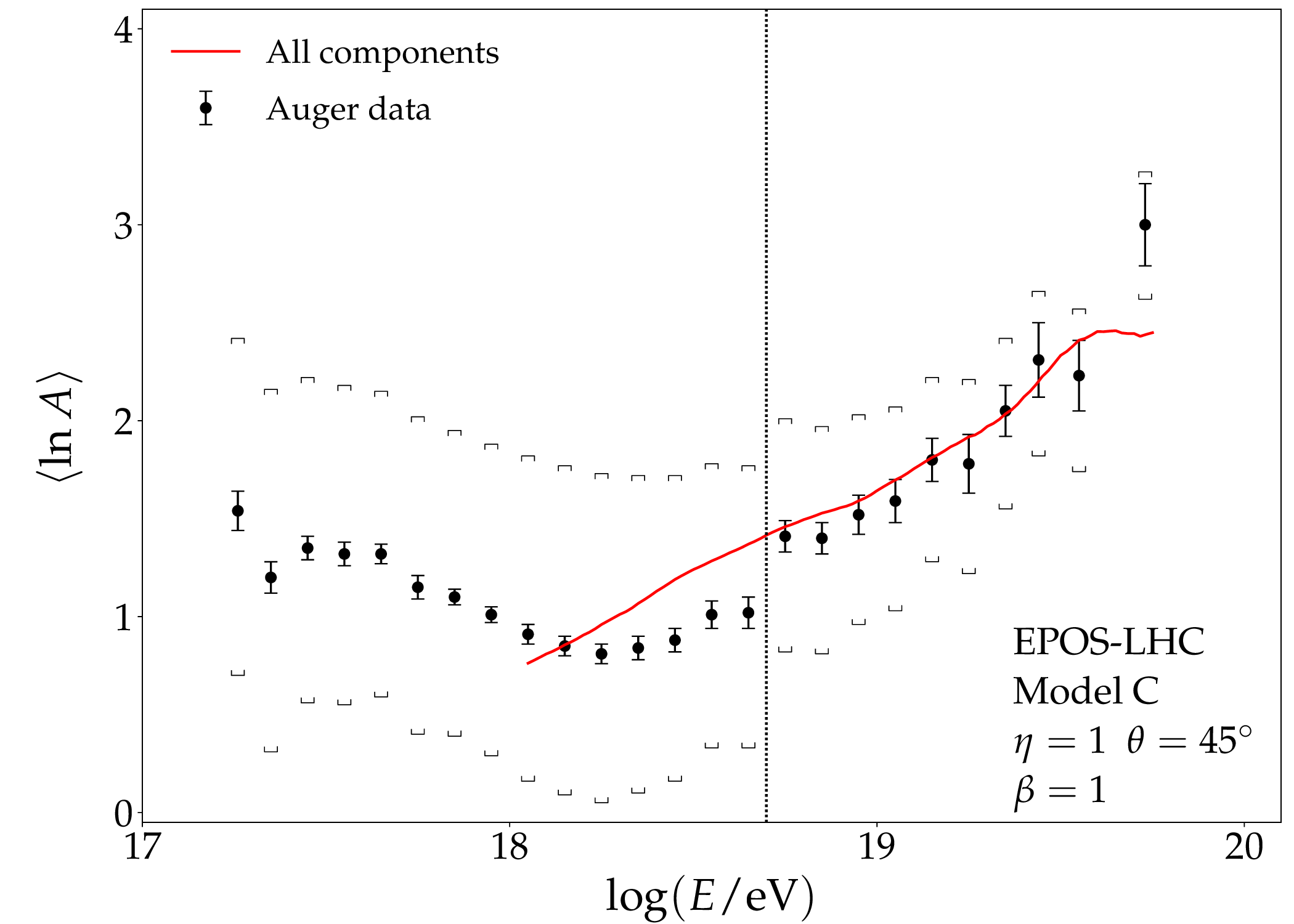}
\includegraphics[width=8cm]{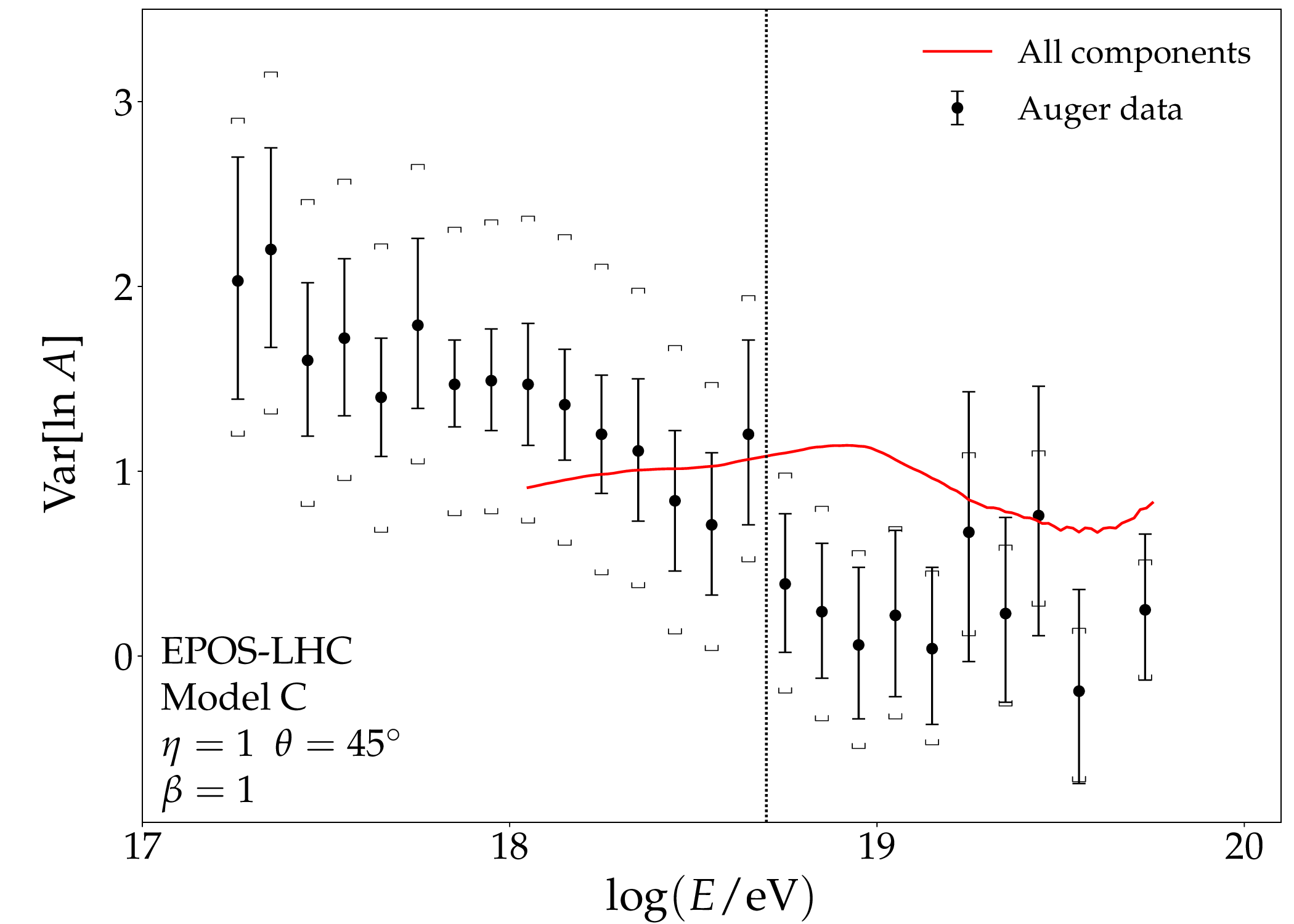}
\caption{Top panel: the cosmic ray flux, multiplied by the cube of the energy, as a function of the logarithm of the 
primary energy. Different cluster contributions to the total flux are shown (see text for details). Middle panel: mean 
value of $\ln A$ as a function of the logarithm of primary energy. Bottom panel: variance of $\ln A$ as a function of 
the logarithm of primary energy. In the three plots the data points represent the Auger measurements and the red solid 
line our best-fit model for $\log(M_\textrm{min}/{\rm M}_\odot)=13$ and $\beta=1$. The vertical lines mark the lower 
energy limit of the data used in the fit. The brackets in the composition data represent the systematic uncertainties.
\label{QParalellB1}}
\end{figure}

While a plasma beta parameter, $\beta=100$, is consistent with current expectations for galaxy cluster accretion shocks, the 
equipartition scenario (a frequently adopted assumption in the literature corresponding to $\beta = 1$) is also considered 
here (see Sec.~\ref{sec:B_acc_sh}). In this case, the maximum energies of the different injected nuclear species become 
larger, thereby generally allowing for a better fit to experimental data, including the flux. Figure~\ref{QParalellB1} shows 
the fit of our model to the Auger data for quasiparallel shocks ($\theta=45^\circ$) in the case of the accretion shock model 
C and high-energy hadronic interaction model EPOS-LHC for $\eta = 1$, $\beta = 1$ and $\log(M_\textrm{min}/{\rm M}_\odot)=13$. 
As in the previous case, the highest energy portion of the flux is dominated by the contribution of the Virgo cluster. Also, 
the contribution from local clusters with virial masses below $10^{14.57}\,$M$_\odot$, and those with larger masses but at 
redshifts $z=[z_0,1]$, are only significant well below the flux suppression energy. The highest energy point of the observed 
UHECR flux is far from the fitted curve, but its error bar is still quite large and then more statistics is required in that 
energy bin. The agreement between the best-fit model and the Auger data composition is quite good. Even though the highest-energy 
data point of the mean value of $\ln A$ appears to deviate from the fit, it remains compatible with it once the total 
uncertainties are taken into account. Similar to the $\beta = 100$ case, the data points for the variance of $\ln A$ in the energy 
range $10^{18.7}$ and $19^{19.2}\,$eV are compatible with the fit within total uncertainties. Table \ref{tab:FitPB1} shows the 
best fit parameters for the case shown in Fig.~\ref{QParalellB1}.
%
%
\begin{table}[!ht]
\caption{Best-fit parameters for the model of Fig.~\ref{QParalellB1}.}.
\vspace{0.2cm}
\centering
\begin{tabular}{lcl}
\hline\hline
$\gamma$        & = & $0.91 \pm 0.03$ \\
$f_\textrm{CR}$ & = & $(1.124 \pm 0.006)\times 10^{-2}$ \\
$I_\textrm{p}$  & = & $0.030 \pm 0.007$ \\
$I_\textrm{He}$ & = & $0.39 \pm 0.01$ \\
$I_\textrm{N}$  & = & $0.51 \pm 0.02$ \\
$I_\textrm{Si}$ & = & $0.07 \pm 0.01$ \\
$I_\textrm{Fe}$ & < & $1.3 \times 10^{-6}$ \\
\hline\hline
\end{tabular}
\label{tab:FitPB1}
\end{table}

\begin{figure}[t]
\centering
\setlength{\abovecaptionskip}{0pt}
\includegraphics[width=7.8cm]{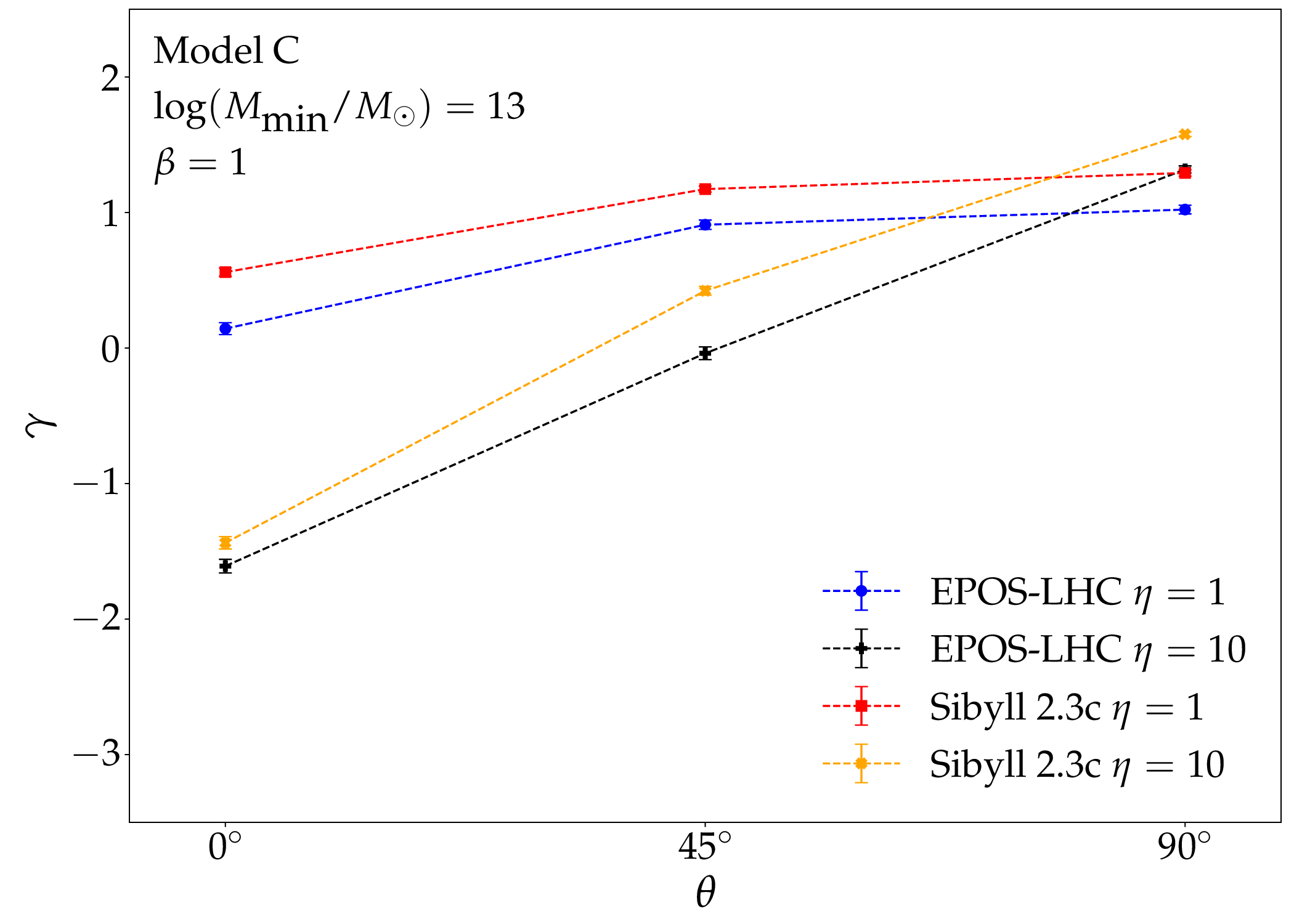}
\includegraphics[width=7.8cm]{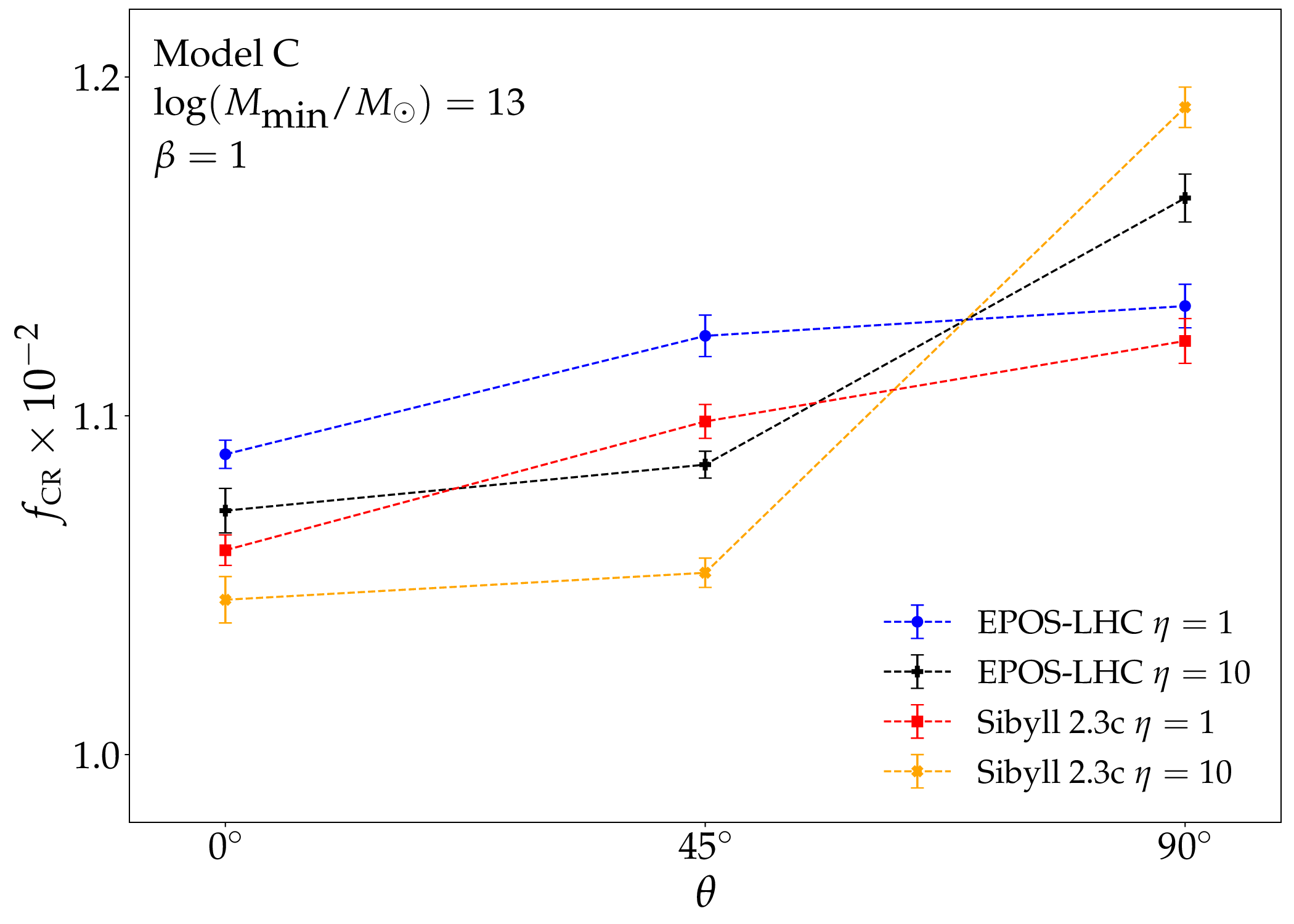}
\caption{Spectral index (top panel) and cosmic ray fraction (bottom panel) as a function of $\theta$ for all cases under 
consideration. The minimum value of the galaxy cluster mass considered here is $\log(M_\textrm{min}/{\rm M}_\odot)=13$
and $\beta=1$.
\label{ParFit}}
\end{figure}

The top panel of Fig.~\ref{ParFit} shows the resulting spectral index of the injected spectrum for the equipartition scenario 
and accretion shock model C as a function of all considered degrees of freedom (i.e., $\eta$, $\theta$ and high-energy hadronic
interaction model). From the figure, it can be seen that $\gamma$ increases with $\theta$, roughly taking values between $-2$ 
and $1.5$. As in the case of $\beta=100$, this is due to the fact that the maximum energy increases with the angle between the 
magnetic field and the shock normal and, then, larger values of $\gamma$ are required for the spectra in order to reach higher 
energies. The bottom panel of Fig.~\ref{ParFit} shows the best-fit values of the CR fraction parameter, $f_\textrm{CR}$, as 
a function of $\theta$. It can be seen from the plot that the average value obtained for the different angles is of the order 
of $0.011$ which is, also in this case, consistent with the results obtained in Ref.~\cite{Inoue:07}. It is worth noting that 
models with $\theta=0$, as well as those with $\theta=45^\circ$ and $\eta=10$, do not reproduce the flux in the suppression 
region. Consequently, their best-fit parameters cannot be assigned a meaningful physical interpretation in terms of the underlying 
accelerated spectrum. 

In contrast, models that successfully fit the suppression region yield positive spectral indexes.
Scenarios with $\theta=45^\circ$ and $\eta=1$ (see Fig.~\ref{QParalellB1} as an example), as well as 
all models with $\theta=90^\circ$, can successfully fit the flux, including the suppression region. The spectral indexes of 
the models that adequately describe the data range from $0.7$ to $1.6$, with the highest values corresponding to $\theta = 90^\circ$ 
and $\eta = 10$. This indicates that models with lower maximum energies have smaller spectral indexes, highlighting their difficulty 
in reproducing the suppression region. In all cases, the fits to the first two moments of $\ln A$ remain compatible with experimental 
data within total uncertainties, regardless of whether the suppression region is accurately modeled.

For model D, which assumes a constant $1\,\mu$G magnetic field across all cluster masses, only the case of $\theta = 90^\circ$ and 
$\eta = 10$ successfully reproduces the suppression region for both high-energy hadronic interaction models considered. The resulting 
spectral indexes fitting the data, including the suppression region, are $\gamma = 1.32 \pm 0.03$ and $1.57 \pm 0.02$ for EPOS-LHC 
and Sibyll 2.3c, respectively. As discussed earlier, in these cases the fitted spectral indexes may be interpreted as unbiased estimates 
of the underlying acceleration spectral index, provided that model D is indeed the correct description of the source environment. Note 
that a scenario with $\eta = 1$ and $\theta = 90^\circ$ also achieves a broad agreement in the suppression region.

Finally, we also investigated the dependence of the fitted parameters on the minimum virial mass of galaxy clusters contributing to 
the UHECR background. Overall, increasing the minimum considered mass to $\log(M_\textrm{min}/{\rm M}_\odot)=13.5$ results in a slight 
decrease in the low-mass component of the flux. This change has a small impact on the fit performance and on the resulting best-fit 
values of the model parameters. Consequently, if $\log(M_\textrm{min}/{\rm M}_\odot)=13.5$, the resulting fitted $\gamma$ and 
$f_\textrm{CR}$ values are slightly larger. This is because lower values of the low-mass component require larger values of 
$f_\textrm{CR}$ to fit the flux at low energies. However, since the other components remain unchanged, larger values of $\gamma$ are 
needed to also fit the highest-energy part of the spectrum.

\section{Discussion}
\label{sec:Disc}

Cosmological simulations show that $\theta$ angles in galaxy cluster accretion shocks can vary considerably \cite{Banfi20}. 
This suggests that the intricate nature of the intracluster medium ensures the presence of suitable acceleration sites for 
CR nuclei in external shocks. Note that our accretion shock models are designed to capture global trends at these locations, 
excluding magnetic field and velocity fluctuations. Therefore, the possibility that small-scale fluctuations along the shock 
surface could create favorable conditions for ion acceleration is not completely ruled out. In particular, larger magnetic 
fields can naturally lead to higher maximum energies. However, if particle acceleration is dominated by only a subset of shock 
patches with favorable conditions for nuclear acceleration, a larger acceleration efficiency, $f_{\rm CR}$, would be required, 
since only a fraction of the total kinetic energy flux is processed through those regions.

On the other hand, particle-in-cell simulations modeling the microphysics of CR acceleration show 
that nuclei are most efficiently accelerated in quasiparallel configurations ($\theta \lesssim 45^{\circ}$) for 
strong shocks \cite{Caprioli14}. However, recent findings show that protons and heavier nuclei might also accelerate 
in quasiperpendicular shocks \cite{Orusa:23}. As stated in Refs.~\cite{Blandford:23,Blandford:25}, preprocessed CR 
populations at galactic scales might be further reaccelerated in galaxy cluster accretion shocks, which could also 
promote more efficient acceleration in quasiperpendicular shocks.   

For $\beta=100$, a significant fraction of the considered parameter space is disfavored by current data. Only models A and C 
(see Table~\ref{tab:models_vsh_B}) with perpendicular shocks and $\eta=10$, within the context of the Sibyll2.3c hadronic model, 
can fit the flux including the suppression region. In the equipartition scenario with $\beta = 1$, which is disfavored by current 
estimates, the UHECR Auger data (both flux and composition) can still be satisfactorily reproduced by models with quasiparallel 
or perpendicular shocks. In general, in all models that reasonably fit the suppression region, the best-fit values of $\gamma$ 
are typically between $0.7$ and $1.6$ (see Figs.~\ref{ParFitB100} and~\ref{ParFit}). This contrasts with the $\gamma = 2$ 
expectation from the first-order Fermi mechanism in the case of strong shocks. Nonetheless, values of $\gamma < 2$, even including 
negative ones, have also been reported in several analyses (see, e.g., Refs.~\cite{AugerFit:17, AugerFit:23}). This suggests that 
the fitted spectral index does not necessarily correspond to the intrinsic acceleration index, as CRs may interact with photon 
and magnetic fields in the acceleration region, modifying the injected spectrum. For accretion shocks in galaxy clusters, an 
additional population of low-energy photons in the outskirts of the shocks can further modify the acceleration spectrum but the 
maximum energy can also be affected. Moreover, the magnetic-horizon effect absent in the present work can introduce a low-energy 
cutoff due to interactions between CRs and the intergalactic magnetic field, and may therefore play an important role. Including 
this effect in the propagation model can reduce the tension between the fitted $\gamma$ values and the expectation from the 
first-order Fermi mechanism, as shown in Ref.~\cite{AugerFitMH:24}. However, the results in Ref.~\cite{AugerFitMH:24} are based 
on a different cutoff shape, source evolution, and modeling framework. Moreover, the inferred $\gamma$ may also reflect the spectral 
index of a hypothetical low-energy component accelerated in the termination shocks of powerful galaxies and subsequently 
reaccelerated in cluster external shocks \cite{Blandford:23, Blandford:25}. In any case, the fact that the spectral indexes 
obtained from fits to Auger data are smaller than those expected from the first-order Fermi mechanism remains as an open problem 
in UHECR physics.

The results obtained for $\beta = 100$ suggest a scenario in which the observed UHECR flux could be composed of a galaxy cluster 
component that dominates at lower energies, together with the contribution of one or a few of other local 
sources--potentially including the radio galaxy Centaurus A--which dominate the flux in the suppression region. In such a scenario, 
it is possible to explore more realistic values of $\gamma$ corresponding to the cluster component, since the physical meaning of a 
parameter in a given model can only be assessed when the model is capable of fully reproducing the experimental data. A model of 
this type will be developed in future studies.

The best-fit values of the CR fraction parameter, $f_\textrm{CR}$, are of order $0.01$ for all models considered, regardless 
of whether they successfully reproduce the suppression region of the flux. This occurs because the fit to $f_\textrm{CR}$ depends 
solely on the flux normalization and is therefore dominated by the flux level at the minimum energy considered, where the 
experimental uncertainties are small. As previously noted, this value is consistent with the estimate reported in 
Ref.~\cite{Inoue:07}, although that analysis adopted a spectral index of $1.7$, which differs significantly from several of the 
values obtained in this work. Moreover, among the models compatible with the experimental data, the minimum $\gamma$ value we 
found is approximately $0.7$. Thus, even if the derived $f_\textrm{CR}$ values are consistent with those reported in 
Ref.~\cite{Inoue:07}, the corresponding accelerated spectra may differ significantly, because models with harder spectra inject 
substantially more particles at high energies, and fewer at lower energies, than models with softer spectra at the same luminosity.

As mentioned before, the highest-energy part of the UHECR flux is dominated by the contribution from Virgo. However, neither 
the Auger nor the TA data shows any excess in the Virgo region (see Sec.~\ref{sec:Int}). This fact could be explained either by the 
inability of cluster accretion shocks to accelerate UHECRs or, even if UHECR acceleration is efficient, by the action of different 
mechanisms, often involving magnetic deflections. In particular, some alternatives have been proposed in this respect. In the model 
of Ref.~\cite{Kim:19}, CRs from the Virgo cluster are able to escape through magnetized filaments before being scattered toward the 
Earth. This model has been developed to explain the TA hotspot, which is located far from the direction of Virgo. Additionally, the 
Galactic magnetic field can play a significant role. In particular, in Ref.~\cite{Allard:24} it is shown that strong magnetic 
demagnification could explain the absence of CR flux excess in this region of the sky.

\section{Conclusions}
\label{sec:Conc}

In this work, we studied the possibility that UHECRs are accelerated in external accretion shocks present in galaxy clusters. We modeled 
the contribution of nearby massive clusters, including Virgo, as a discrete set, while the remaining sources were assumed to follow a 
continuous distribution given by the number density of cluster halos which, in turn, is determined by the redshift-dependent HMF. Assuming 
that sources inject multiple nuclear species with a spectrum following a power law with an exponential cutoff, we performed fits both for 
the flux at Earth and the composition profile measured by the Pierre Auger Observatory. We found that the flux at the highest energies is 
dominated by Virgo, while the contribution of the remaining nearby clusters is less significant, mainly involving smaller energies. 

After considering a series of accretion shock models with plasma beta parameter $\beta=100$ (see Table~\ref{tab:models_vsh_B}), i.e. 
consistent with expectations in galaxy cluster outskirts, we found that all models with $\theta\sim 45^{\circ}$ can reasonably reproduce 
both the mean and variance of the CR composition within total uncertainties. However, they do not properly fit the ultrahigh-energy 
suppression of the CR flux, thus failing to reproduce the highest-energy UHECRs that originate from direct acceleration of thermal 
particles. Only some models with $\eta=10$ and $\theta=90^{\circ}$ can additionally fit the flux in the suppression region. This suggests 
that additional, prior reacceleration may be necessary to reproduce the data within the framework of these accretion shock scenarios. 
For accretion shock models with $\beta=1$, which are disfavored by current theoretical expectations, the resulting magnetic fields at 
the shock locations are roughly an order of magnitude stronger than in the $\beta=100$ case. Consequently, both the composition and 
flux of the UHECR spectrum can also be reasonably reproduced, within uncertainties. 

The best-fit spectral indexes for the models that successfully describe the experimental data lie between approximately $0.7$ and $1.6$, 
in agreement with other analyses based on Pierre Auger Observatory data. Concerning the fraction of kinetic energy of the infalling material 
that must be converted into CRs, we find that all models consistent with the data require efficiencies of a few percent, also compatible 
with previous results.

The absence of an excess in the CR distribution toward the direction of Virgo may be attributed either to the inability of accretion 
shocks to accelerate nuclei to the highest observed energies or, potentially, for scenarios reproducing the suppression region, to the 
interaction of the accelerated CRs with the magnetic fields of the filaments connected to Virgo, and/or to their deflection within the 
Galactic magnetic field.
 
Finally, based on our results, we conclude that accretion shocks in galaxy clusters can contribute significantly to the origin of the 
most energetic UHECRs only if magnetic fields are stronger than typically expected in these environments, which could arise from the 
existence of local fluctuations in the shock front properties, or in the case of perpendicular shock normal-magnetic field configurations. 
This scenario, however, may also account for all or part of the observed flux at intermediate energies.

\begin{acknowledgments}
A.D.S. and S.E.N. are members of the Carrera del Investigador Cient\'{\i}fico of CONICET, Argentina. 
They acknowledge support from CONICET (PIBAA R73734 and PIP 11220200102979CO). S.~E.~N. also acknowledges support from Agencia Nacional de Promoci\'on Cient\'{\i}fica y Tecnol\'ogica (PICT 2021-GRF-TI-00290). 
\end{acknowledgments}

\section{DATA AVAILABILITY}

The data that support the findings of this article are not publicly available. The data are available from the authors
upon reasonable request.

\appendix

\section{\label{ApIntL} CONTRIBUTION OF DIFFERENT PROCESSES TO THE TOTAL MEAN FREE PATH}

\begin{figure}[t!]
\vspace{1cm}
\centering
\setlength{\abovecaptionskip}{0pt}
\includegraphics[width=7.8cm]{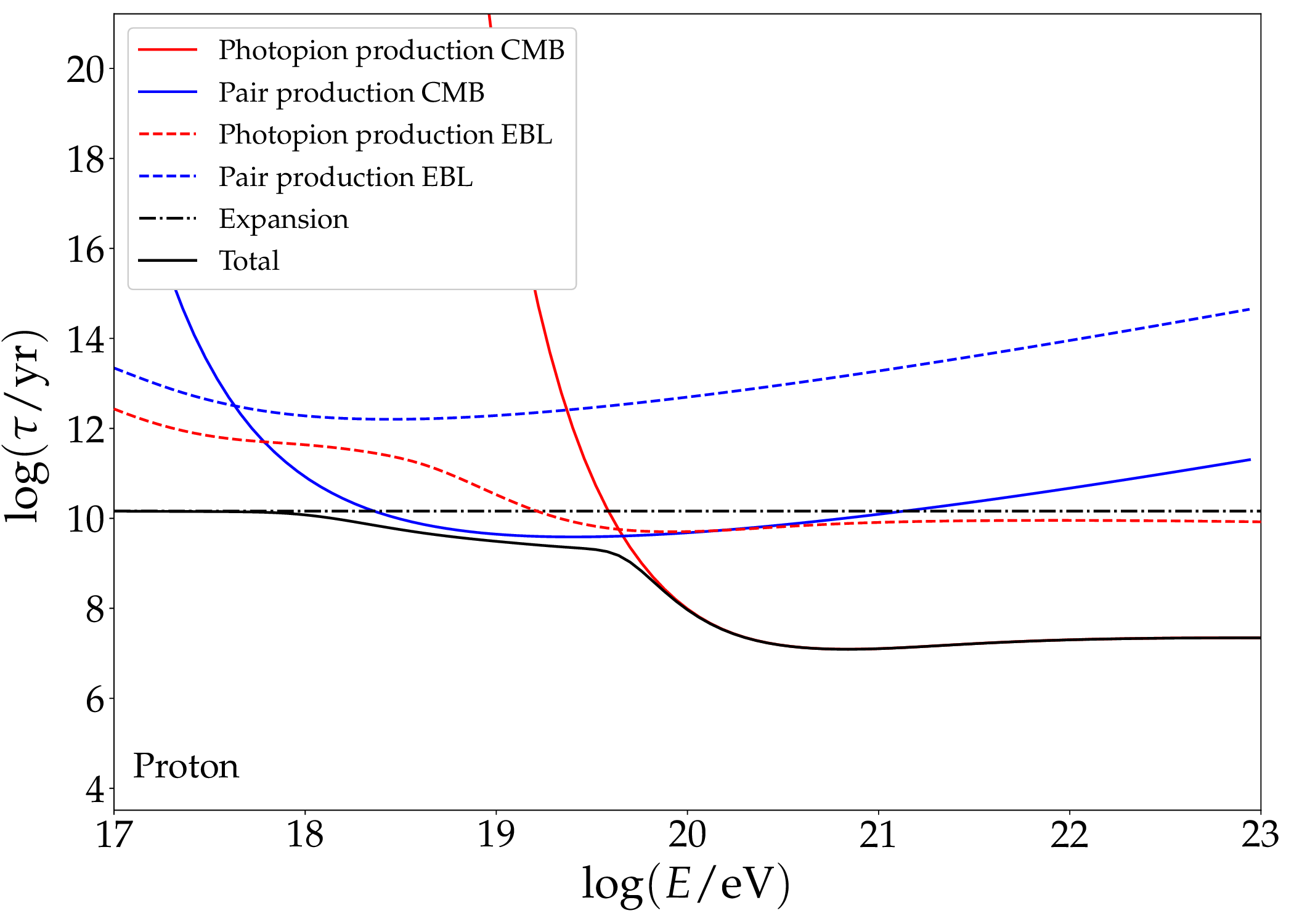}
\includegraphics[width=7.8cm]{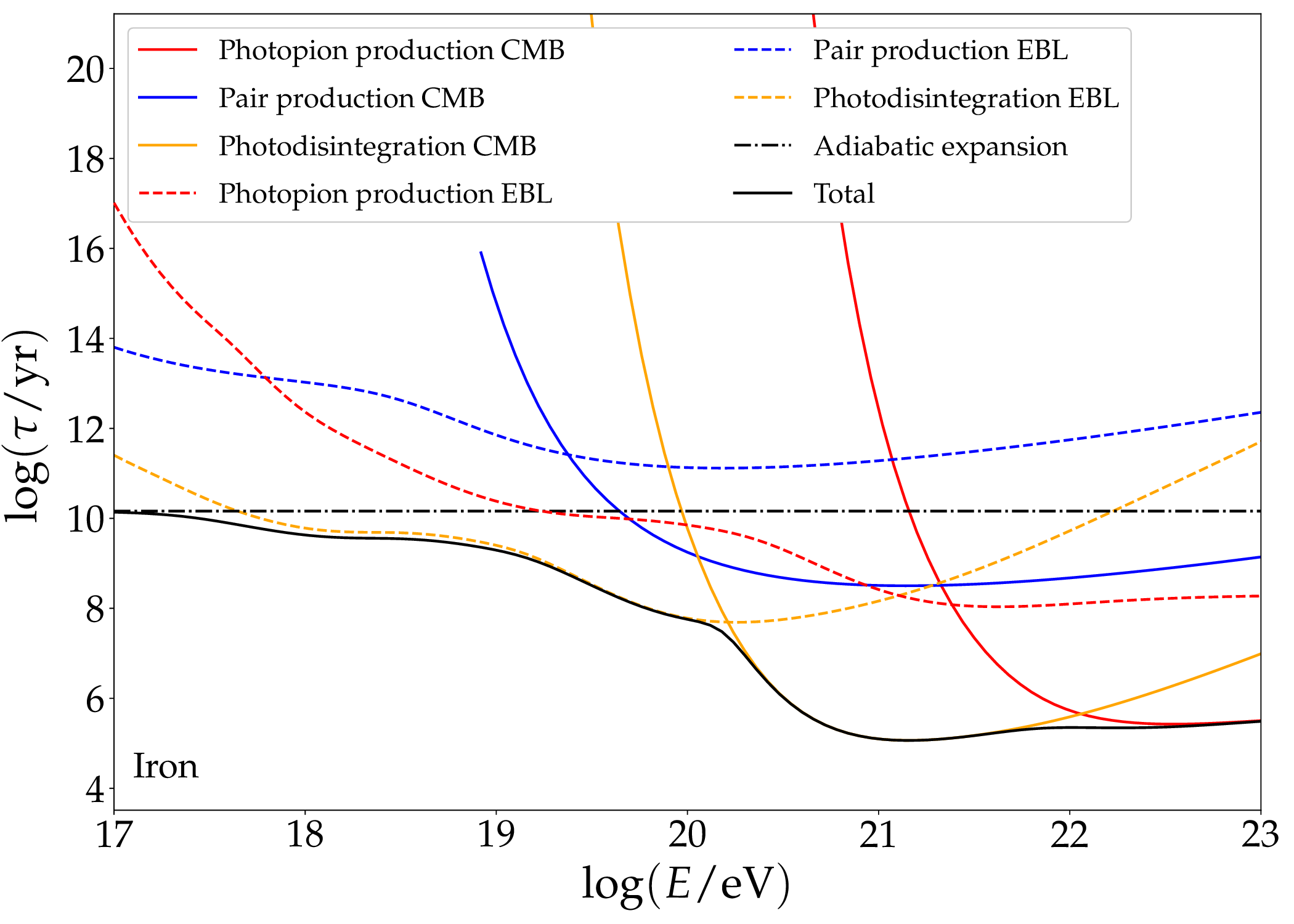}
\caption{Logarithm of the mean interaction time as a function of the logarithm of energy for protons 
(top panel) and iron nuclei (bottom panel) at redshift zero. \label{MaxEnLint}}
\end{figure}

As mentioned in Sec.~\ref{sec:MaxEn}, the processes included in the interaction time calculation 
are: photo-pion production, photo-disintegration, and pair production. The energy loss due to the 
adiabatic expansion of the Universe is also considered. Figure~\ref{MaxEnLint} shows the logarithm of 
the mean interaction time of proton and iron nuclei as a function of the logarithm of energy including 
the contribution of each individual process to the total interaction time. The photon fields considered 
are the CMB and the EBL. The EBL model used in the calculations is that of Ref.~\cite{Gilmore:12}.   

\section{\label{ApModelCSib} FIT TO ACCRETION SHOCK MODEL C}

\begin{figure}[t!]
\centering
\setlength{\abovecaptionskip}{0pt}
\includegraphics[width=8cm]{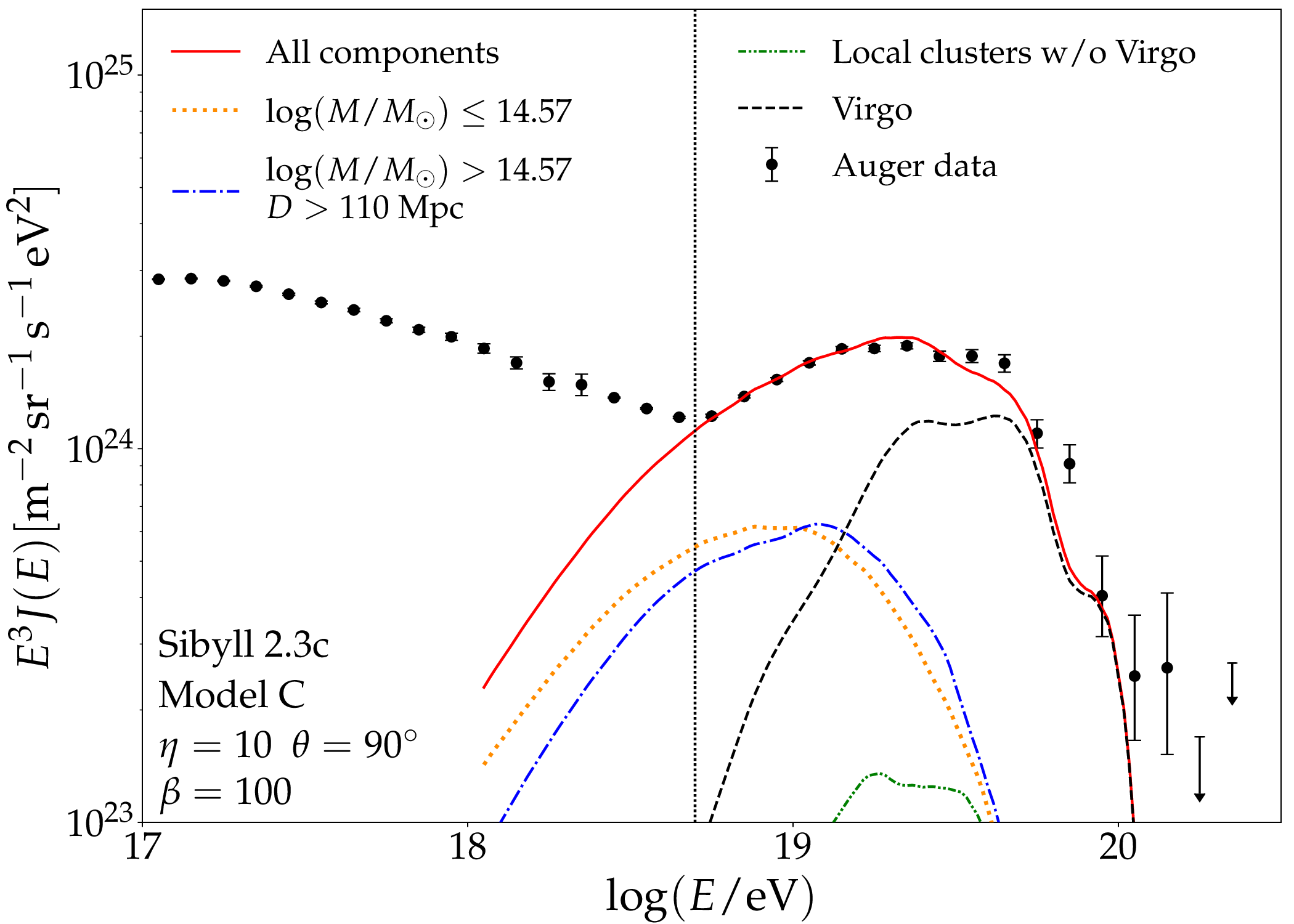}
\includegraphics[width=8cm]{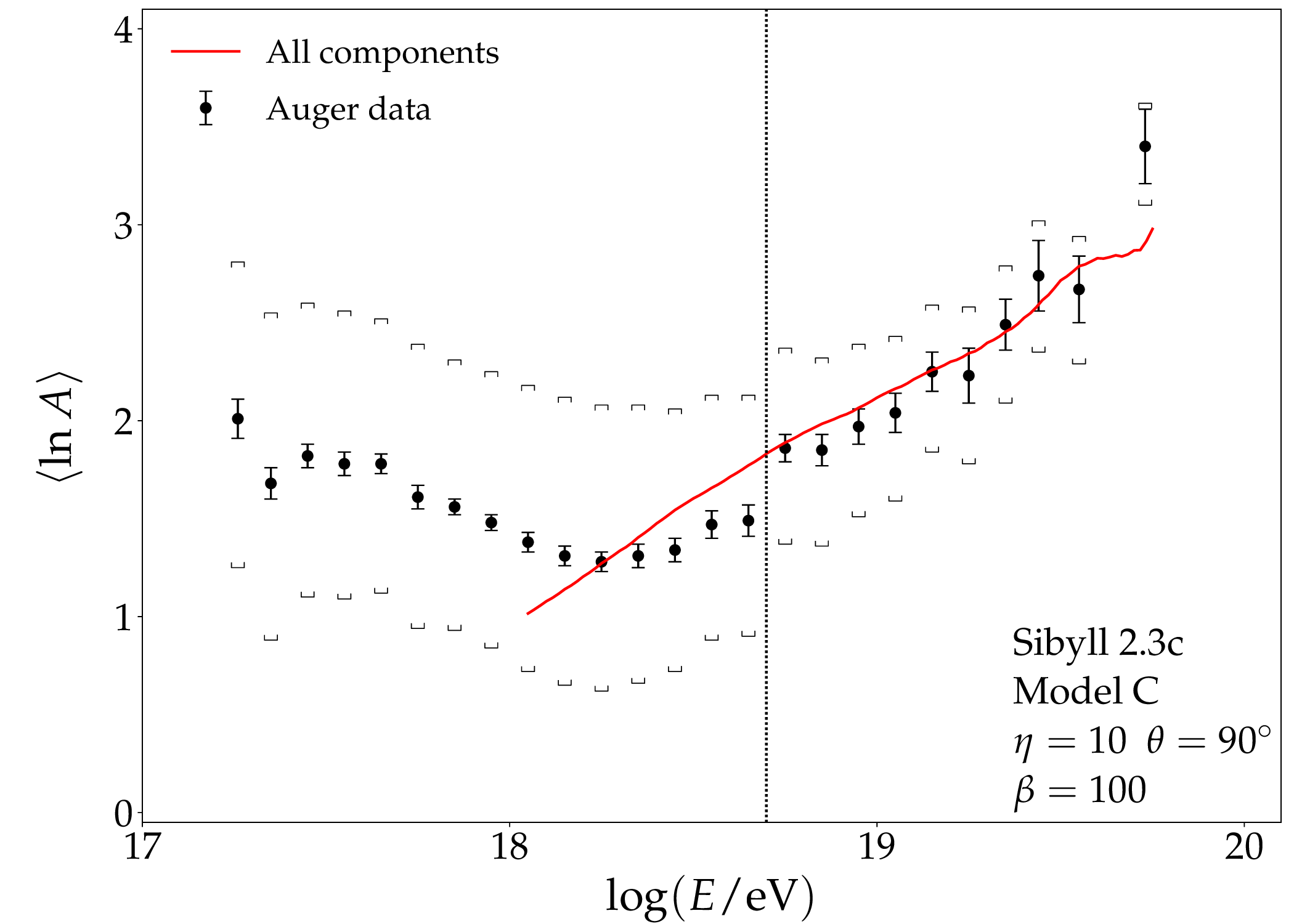}
\includegraphics[width=8cm]{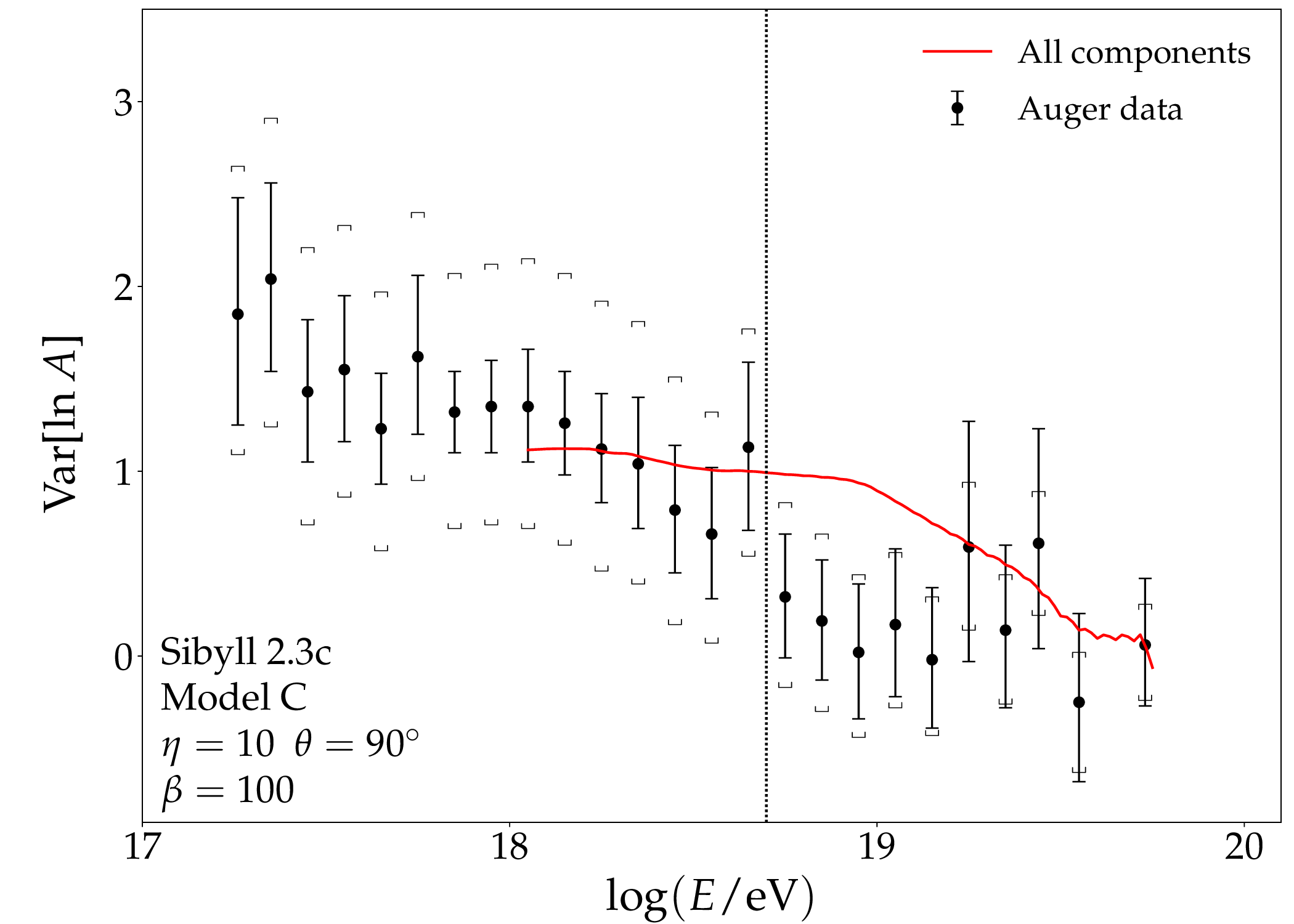}
\caption{Top panel: the cosmic ray flux, multiplied by the cube of the energy, as a function of the logarithm of the 
primary energy. Different cluster contributions to the total flux are shown. Middle panel: mean value of $\ln A$ as 
a function of the logarithm of primary energy. Bottom panel: variance of $\ln A$ as a function of the logarithm of 
primary energy. In the three plots the data points represent the Auger measurements and the red solid line our 
best-fit model for $\log(M_\textrm{min}/{\rm M}_\odot)=13$ and $\beta=100$. The vertical lines mark the lower energy 
limit of the data used in the fit. The brackets in the composition data represent the systematic uncertainties.
\label{PerpendicularCSib}}
\end{figure}

Figure \ref{PerpendicularCSib} presents the fit of both the flux and composition data for the accretion shock model C 
with $\beta = 100$ for perpendicular shocks, $\eta=10$, and Sibyll2.1c as the high-energy hadronic interaction model. 
This model provides a reasonable fit to the composition data and can better reproduce the suppression region of the flux. 
Additionally, Table \ref{tab:FitMCapp} shows the corresponding best-fit parameters. Note that similar results are obtained 
for the accretion shock model A with $\beta = 100$ for perpendicular shocks, $\eta=10$, and Sibyll2.1c.
%
%
\begin{table}[t!]
\caption{Best-fit parameters for the model of Fig.~\ref{PerpendicularCSib}.}.
\vspace{0.2cm}
\centering
\begin{tabular}{lcl}
\hline\hline
$\gamma$        & = & $ 1.57 \pm 0.02$ \\
$f_\textrm{CR}$ & = & $( 1.072 \pm 0.007)\times 10^{-2}$ \\
$I_\textrm{p}$  & < & $ 2\times10^{-6}$ \\
$I_\textrm{He}$ & = & $ 0.21 \pm 0.02$ \\
$I_\textrm{N}$  & = & $ 0.62 \pm 0.02$ \\
$I_\textrm{Si}$ & = & $ 0.15\pm 0.01$ \\
$I_\textrm{Fe}$ & = & $ 0.02 \pm 0.01$ \\
\hline\hline
\end{tabular}
\label{tab:FitMCapp}
\end{table}
\clearpage

\bibliography{main}

\end{document}